%% file: main.tex
\pdfoutput=1
\documentclass[preprint,3p]{elsarticle}

\input{preamble.tex}

\usepackage{amssymb}
\usepackage{amsmath}
\numberwithin{equation}{section}

\usepackage[colorlinks=true, linkcolor=blue, citecolor=red, urlcolor=blue]{hyperref}

\usepackage{enumitem}
\usepackage{layouts}
\usepackage{pgfplots}
\pgfplotsset{compat=newest}

\journal{Journal of Computational Physics}

\begin{document}

\begin{frontmatter}



\title{Structure-Preserving MHD-Driftkinetic Discretization for Wave-Particle Interactions}

\author[1,2]{Byung Kyu Na}
\author[1]{Stefan Possanner}
\author[1]{Xin Wang}

\affiliation[1]{organization={Max Planck Institute for Plasma Physics},%Department and Organization
            addressline={Boltzmannstrasse 2}, 
            city={Garching},
            postcode={85748}, 
            state={Bayern},
            country={Germany}}
\affiliation[2]{organization={Technical University of Munich},%Department and Organization
            addressline={Boltzmannstrasse 3}, 
            city={Garching},
            postcode={85748}, 
            state={Bayern},
            country={Germany}}

%% Abstract
\begin{abstract}
We present a structure-preserving discretization of the hybrid magnetohydrodynamics (MHD)-driftkinetic system for simulations of low-frequency wave-particle interactions. The model equations are derived from a variational principle, assuring energetically consistent couplings between MHD fluids and driftkinetic particles. The spatial discretization is based on a finite-element-exterior-calculus (FEEC) framework for the MHD and a particle-in-cell (PIC) method for the driftkinetic. A key feature of the scheme is the inclusion of the non-quadratic particle magnetic moment energy term in the Hamiltonian, which is introduced by the guiding-center approximation. The resulting discrete Hamiltonian structure naturally organizes the dynamics into skew-symmetric subsystems, enabling balanced energy exchange. To handle the non-quadratic energy term, we develop energy-preserving time integrators based on discrete gradient methods. The algorithm is implemented in the open-source Python package \texttt{STRUPHY}. Numerical experiments confirm the energy-conserving property of the scheme and demonstrate the capability to simulate energetic particles (EP) induced excitation of toroidal Alfvén eigenmodes (TAE) without artificial dissipation or mode filtering. This capability highlights the potential of structure-preserving schemes for high-fidelity simulations of hybrid systems.
\end{abstract}


%\item Structure-preserving discretization of a hybrid MHD-driftkinetic model.
%\item Energy-preserving integrators based on discrete gradient methods.
%\item Simulations of wave-particle interactions in toroidal geometry without artificial dissipation or %Fourier mode filtering.
%\item An open source Python package \texttt{STRUPHY} providing reproducible simulations and diagnostics.

%% Keywords

\end{frontmatter}

\section{Introduction}\label{sec1}
Plasma dynamics is inherently multiscale, including magnetohydrodynamics (MHD) waves, macroscopic flows and microscopic particle motions, which mutually interact across widely separated temporal and spatial scales. A notable example is wave-particle interaction, where waves exchange energy with resonant particles, e.g., Landau damping \cite{Chen_2016}. In fusion plasmas, such interactions occur when energetic particles (EP), produced by auxiliary heating or fusion reactions, resonate with toroidal Alfvén eigenmodes (TAE) and thus degrade plasma stability and confinement significantly \cite{Vlad_1999, Heidbrink_2008, Todo_2019}. Similar processes, for instance, interactions between the solar wind and the magnetosphere \cite{Lysak_2023}, are observed in astrophysical plasmas. To capture these processes, hybrid MHD-kinetic models have been developed, where bulk plasma is modeled as MHD fluids, while EPs are treated kinetically as a separate species since EPs are typically far from thermal equilibrium. Their basic formulation is obtained by incorporating the current or pressure contributions of the kinetic species---given as moments of the kinetic distribution---into the MHD momentum balance equation \cite{Park_1992}. In this way, hybrid models provide an intermediate description, including the kinetic effects of EPs while being computationally feasible on macroscopic scales. Hybrid modeling itself, however, is an open and interesting topic: across existing codes, different coupling schemes and various kinetic reductions are employed, and the implications of these choices---both from a modeling and computational perspective---are not yet fully understood. A particular concern is the lack of Hamiltonian structure, so exact energy conservation may be lost, which in turn can lead to unphysical instabilities \cite{Tronci_2014}, and necessitate artificial damping or filtering for numerical stability. One way to address this problem is to derive hybrid models from variational principles \cite{Holm_2012, Burby_2017, Close_2018, Tronci_2020}, which guarantee exact energy conservation while at the same time providing a systematic framework for a better understanding of hybrid modeling.

Along with the theoretical developments, significant progress has been made on the numerical side with the advent of structure-preserving discretizations, refer to \cite{Morrison_2017, Razafindralandy_2018} for a review. The methods aim to retain the geometric structures of the continuous system at the discrete level, thereby ensuring conservation laws and long-term stability of simulations. The basic strategy is to discretize directly the Poisson bracket or variational structure; this approach has been extensively developed for Vlasov–Maxwell systems (e.g., \cite{He_2015, Qin_2016, Kraus_2017}). An alternative strategy is to preserve the skew-symmetric structure of the discrete Poisson matrix and employ integral-preserving implicit time integrators such as Crank-Nicolson \cite{Crank_Nicolson_1947} and discrete gradient \cite{McLachlan_1999} methods. The strategy has been successfully applied to Vlasov–Ampère \cite{Chen_2011} and Vlasov–Maxwell \cite{Kormann_2021} systems. Moreover, the framework was also extended to a hybrid MHD–kinetic model with a full-orbit Vlasov description of the kinetic species~\cite{Holderied_2021}. Nevertheless, the application of these models to realistic fusion simulations has so far remained limited, primarily due to the short time step imposed by the full-orbit description. This motivates the present work, which extends the framework to MHD-driftkinetic systems. The driftkinetic reduction enables the use of much larger time steps, better adapted to the time scale of TAEs. A tailored discretization is introduced to incorporate the particle magnetic moment energy, which appears in the Hamiltonian of the MHD-driftkinetic system through the guiding-center approximation. The energy term is non-quadratic, depending simultaneously on the magnetic field and the kinetic distribution, and leads to unique couplings that naturally capture guiding-center magnetization in Alfvénic dynamics and consistently link particle drifts with the evolution of the MHD flow velocity. To ensure exact conservation of this non-quadratic energy, we develop energy-preserving time integrators based on discrete gradient methods. The scheme is implemented in the open-source Python package \texttt{STRUPHY} \cite{Stefan_2023, STRUPHY}, which features a finite-element-exterior-calculus (FEEC) approach for MHD equations with a particle-in-cell (PIC) treatment of the kinetic species and provides a range of diagnostics for wave-particle interactions.
 
The remainder of this article is organized as follows. \autoref{sec2} introduces the model equation together with the selective linearization strategy adopted in this work. \autoref{sec3} presents the basic concept of the spatial discretization within the structure-preserving framework and establishes the Hamiltonian structure of the resulting semi-discrete system. In \autoref{sec4}, we describe the time discretization strategy based on Poisson splitting combined with energy-conserving integrators. \autoref{sec5} reports two numerical experiments; one in a periodic slab configuration to verify exact energy conservation of the scheme, and another in toroidal geometry to demonstrate its capability on the TAE benchmark case \cite{Könies_2018}. Finally, \autoref{sec6} summarizes the work and provides an outlook. 
 
\section{Model equations}\label{sec2}
\subsection{Hamiltonian hybrid MHD-driftkinetic with current-coupling scheme}\label{subsec2-1}
The model equation of the present work is based on the hybrid MHD-driftkinetic model derived from the variational mean-fluctuation splitting approach, as presented in Section 6.2 of \cite{Tronci_2020}:
\begin{equation} \label{varprinc}
\begin{aligned}
    \delta \int^{t_2}_{t_1} \left[ \int \fh \left( (\mh \vp \bb_0 + \qh \Ab)\cdot \ub_\text{gc} - \frac{\mh}{2}\vp^2 - \mu (\bb_0 \cdot \Bb) - \qh \Ub \cdot \Ab \right) \dx \dvp \dmu \right.
    \\
    \left. +\int \left(\frac{\rho}{2}|\Ub|^2 - \frac{p}{\gamma - 1} - \frac{1}{2\mu_0} |\Bb|^2\right) \dx \right] \tn{d}t = 0 \,.
\end{aligned}
\end{equation}
Here, the variables $\rho$, $\Ub$ and $p$ represent the mass density, flow velocity and pressure of the MHD fluids with adiabatic exponent $\gamma = 5/3$, respectively. The symbol $\fh$ refers to a driftkinetic distribution function in $(\xb, \vp, \mu)$-phase-space, modeling energetic ("hot") particles with mass $\mh$ and charge $\qh$. Moreover, $\bb_0$ stands for the unit vector of a fixed magnetic background field, comprised within the total magnetic field $\Bb = \nabla \times \Ab$, and $\ub_\text{gc} = \dot \xb$ denotes the guiding-center velocity.
The variational principle \eqref{varprinc} leads to the equations
\begin{subequations}
\begin{align}
    &\pdt \rho + \nabla \cdot (\rho \Ub) = 0 \,, \label{hybrid_continuity}
    \\
    \rho &\pdt \Ub + \rho (\Ub \cdot \nabla) \Ub - \left( \frac{1}{\mu_0} \nabla \times \Bb + \qh \ngc \Ub - \Jb_{\textrm{gc}} - \nabla \times \Mb_{\textrm{gc}} \right) \times \Bb + \nabla p = 0\,, \label{hybrid_momentumbalance}
    \\
    &\pdt \Bb + \nabla \times (\Bb \times \Ub) = 0 \,, \label{hybrid_induction}
    \\
    &\pdt p + \nabla \cdot (p \Ub) + (\gamma - 1) p \nabla \cdot \Ub = 0 \,, \, \label{hybrid_pressure}
    \\
    &\pdt \fh + \pder{}{\xb} \cdot \left[ \frac{\fh}{\Bpa} (\vp \Bb^* - \bb_0 \times \Eb^*)\right] + \frac{\qh}{\mh} \pder{}{\vp} \left( \frac{\fh}{\Bpa} \Bb^* \cdot \Eb^*\right) = 0 \,, \, \label{hybrid_driftkinetic}
\end{align}    
\end{subequations}
where the three moments of the kinetic distribution in \eqref{hybrid_momentumbalance} are defined as follows:
\begin{subequations}
\begin{align}
    &\ngc(\fh) = \int \fh \dvp \dmu \,, 
    \\
    &\Jgc(\fh,\Bb,\Ub) = \qh \int \frac{\fh}{\Bpa} (\vp \Bb^* - \bb_0 \times \Eb^*) \dvp \dmu \,, \,
    \\
    &\Mgc(\fh) = - \int \fh \mu \bb_0 \dvp \dmu \,, 
\end{align}
\end{subequations}
with the effective electromagnetic fields $\Bb^* = \Bb + \mh/\qh \vp \nabla \times \bb_0$ and $\Eb^* = - \Ub \times \Bb - \mu/\qh \nabla (\bb_0 \cdot \Bb)$ in guiding-center approximation and the parallel effective magnetic field $\Bpa = \Bb^* \cdot \bb_0$. 

The model \eqref{hybrid_continuity}-\eqref{hybrid_driftkinetic} features the ideal MHD equations, self-consistently coupled to the driftkinetic equation for $\fh$ via the so-called "current-coupling" scheme. In this scheme, the EPs act on the fluid flow through a modification of the MHD current $\Jb = \nabla \times \Bb / \mu_0$, represented by the three terms with "gc" in equation \eqref{hybrid_momentumbalance}. 
One of the main features of the model is the mean-fluctuation splitting approach, which leads to simplified energy-conserving couplings. The key idea is to apply gyro-averaging along a time-independent background magnetic field $\bb_0$. In general approaches, the gyroradius is considered to be perpendicular to the total magnetic field $\bb = \Bb/|\Bb|$. Consequently, the system avoids the analytically and numerically complex terms which must be introduced to construct energy-conserving couplings between MHD and kinetic equations with guiding-center approximation (see \cite{Burby_2017} for details) while preserving the Hamiltonian, which is the total energy of the system:
\begin{equation}\label{Hamiltonian0}
    \mathcal{H} = \int \frac{\rho}{2} |\Ub|^2 \dx + \int \frac{1}{2\mu} |\Bb|^2 \dx + \int \frac{p}{\gamma -1} \dx + \int \fh \frac{\mh}{2}\vp^2\dvp \dmu \dx + \int \fh \mu (\bb_0 \cdot \Bb)\dvp \dmu \dx \,.
\end{equation}
In summary, the choice of the model is motivated by two key properties:
\begin{enumerate}
    \item It is derived from a variational principle, ensuring energy-conserving coupling between bulk plasma and kinetic particles, which is closely linked to the numerical stability of hybrid simulations;
    \item It avoids numerically challenging terms, enabling the construction of a computationally efficient scheme suitable for practical use in physics research.
\end{enumerate}

\subsection{Linearization}\label{subsec2-2}
In this study, we adopt a selective linearization strategy for the MHD equations \eqref{hybrid_continuity}-\eqref{hybrid_pressure}, wherein nonlinearities involving density and pressure fluctuations are neglected, while the magnetic field-to-flow coupling responsible for shear Alfvén dynamics remains in its fully nonlinear form. Meanwhile, the kinetic equation \eqref{hybrid_driftkinetic} and the MHD-kinetic coupling terms in the momentum balance equation \eqref{hybrid_momentumbalance} remain unmodified, thereby preserving the exact energy-conserving couplings. With this approach, the model includes nonlinear Alfvén dynamics and their interactions with particles, while simplifying the treatment of compressive effects such as slow and fast magnetosonic waves. Furthermore, the global MHD equilibrium dynamics can be excluded ("$\delta f$-approach" for MHD), allowing a focused investigation of wave-particle interactions.

\subsubsection{Linearization scheme 1}
As a first linearization scheme, we linearize the system on the zero-flow MHD equilibrium ($\rho_0$, $\Ub_0=0$, $\Bb_0$, $p_0$) such that
\begin{equation}\label{equilibrium_1}
    \Jb_\mr{0} \times \Bb_0 - \nabla p_0 = 0 \,,
\end{equation}
where $\Jb_\mr{0} = 1/\mu_0 (\nabla \times \Bb_0)$ is the MHD equilibrium current density. On this equilibrium, the first linearized system can be obtained by applying the general perturbation ansatzes $(\rho = \rho_0 + \wt \rho,\, \Ub = \wt \Ub,\, \Bb = \Bb_0 + \wt \Bb,\, p = p_0 + \wt p)$ and retaining the nonlinear magnetic field-to-flow coupling terms:
\begin{equation}\label{linear_system_1}
    \textnormal{Scheme 1 }\left\{
    \begin{aligned}
    \pdt{\wt \rho} &+ \nabla \cdot (\rho_0 \wt \Ub) = 0 \,, 
    \\
    \rho_0 \pdt{\wt \Ub} &- \Jb_0 \times \wt \Bb -  \frac{1}{\mu_0} (\nabla \times \wt \Bb) \times \Bb + \nabla \wt p 
    \\
    &- \left(\qh \ngc \wt \Ub -\Jgc(\fh,\Bb,\wt \Ub) - \nabla \times \Mgc(\fh) \right) \times \Bb = 0 \,, 
    \\
    \pdt{\wt \Bb} &+ \nabla \times (\Bb \times \wt \Ub) = 0 \,,
    \\
    \pdt{\wt p} &+ \nabla \cdot (p_0 \wt \Ub) + (\gamma - 1) p_0 \nabla \cdot \wt \Ub = 0 \,.
    \end{aligned}\right.
\end{equation}
Remark in particular the terms featuring the full magnetic field $\Bb$, which are nonlinear. The driftkinetic equation \eqref{hybrid_driftkinetic} remains unchanged.
Due to this partial linearization, the Hamiltonian \eqref{Hamiltonian0} is no longer conserved and the perturbed Hamiltonian
\begin{equation}\label{pHamiltonian1}
    \wt{\mathcal{H}}_1(t) = \int \frac{\rho_0}{2} |\wt \Ub|^2 \dx + \int \frac{1}{2\mu_0} |\wt \Bb|^2 \dx + \int \frac{\tilde p}{\gamma -1} \dx + \int \fh \frac{\mh}{2}\vp^2\dvp \dmu \dx + \int \fh \mu (\bb_0 \cdot \Bb)\dvp \dmu \dx  \,,
\end{equation}
is evolving in time as
\begin{equation}
    \dt{\wt{\mathcal{H}}_1} = \int(\nabla \times \Bb_0) \times \wt \Bb \cdot \wt \Ub - (p_0 - \tilde p)\nabla \cdot \wt \Ub \dx \,.
\end{equation}

\subsubsection{Linearization scheme 2}
The linearization approach used in \eqref{linear_system_1} is identical to the approach used in \cite{Holderied_2021}, where the MHD equations are coupled with the particle distribution evolving with the full-orbit Vlasov equations
\begin{equation}\label{MHD-Vlasov}
    \textnormal{MHD-Vlasov }\left\{
\begin{aligned}
    &\rho \pdt{\Ub} + \rho(\Ub \cdot \nabla)\Ub - \left(\frac{1}{\mu_0}\nabla \times \Bb + \qh n_\textnormal{h} \Ub -\Jb_\textnormal{h}(\fh) \right) \times \Bb + \nabla p = 0\,,
    \\
    &n_\mr{h}(\fh) = \int \fh \tn d^3 v \,, \qquad \Jb_\mr{h}(\fh) = \qh\int \fh \vb \tn d^3 v \,.
\end{aligned} \right.
\end{equation}
In case of the MHD-Vlasov system \eqref{MHD-Vlasov}, assuming that a) the general MHD equilibrium condition \eqref{equilibrium_1} holds and b) the hot particles have only a parallel equilibrium current $\Jb_\mr{h0} = J_{\parallel} \bb_0$, the zeroth-order terms give
\begin{equation} \label{vlasov:equil}
    \left[\frac{1}{\mu_0}\nabla \times \Bb_0 - J_{\parallel} \bb_0 \right] \times \Bb_0 - \nabla p_0 = 0 \,.
\end{equation}
However, when it comes to MHD-driftkinetic system \eqref{hybrid_continuity}-\eqref{hybrid_driftkinetic}, assuming again that the general MHD equilibrium condition \eqref{equilibrium_1} holds, there zeroth-order terms in \eqref{hybrid_momentumbalance} yield
\begin{equation}
    \left[\frac{1}{\mu_0}\nabla \times \Bb_0  -  \underbrace{\Jb_{\textrm{gc},0} - \nabla \times \Mb_{\textrm{gc},0}}_{\neq 0} \right]\times \Bb_0- \nabla p_0 \neq 0 \,.
\end{equation}
Let us prove that indeed a nonzero contribution remains from this term.
If we take the limit $B^*_\parallel \to B_0$ for simplicity, the sum of the two kinetic terms reads
\begin{equation}\label{sum_of_two_moments}
    \begin{aligned}
        \Jb_{\textrm{gc},0} + \nabla \times \Mb_{\textrm{gc},0} &= \int \frac{\fh}{B_0}\left[\qh\vp \Bb_0 + \mh\vp^2 (\nabla \times \bb_0) + \mu\bb_0 \times \nabla B_0 \right]\dvp \dmu - \nabla \times \int \fh \mu \bb_0 \dvp \dmu 
        \\
        &= J_{\parallel}\bb_0 + \frac{p_\parallel}{B_0} (\nabla \times \bb_0) + \frac{p_\perp}{B_0} \bb_0 \times \frac{\nabla B_0}{B_0} - \nabla \times \left( \frac{p_\perp}{B_0} \bb_0 \right) 
        \\
        &= J_{\parallel}\bb_0 + \frac{p_\parallel - p_\perp}{B} (\nabla \times \bb_0) - \frac{\nabla p_\perp}{B_0} \times \bb_0 \,,
    \end{aligned}
\end{equation}
where the parallel and perpendicular pressures are defined as $p_\parallel = \mh \int \fh \vp^2 \dvp \dmu$ and $p_\perp = \int \fh \mu B_0 \dvp \dmu$, respectively. Compared to the current in \eqref{vlasov:equil}, there are two additional terms in the balance equation, which in general will not cancel out. For instance, even in the case of an isotropic velocity distribution with $p_\parallel = p_\perp =: p_\textnormal{h}$, a diamagnetic current remains:
\begin{equation}
    \Bb_0 \times \left(\Jb_{\textrm{gc},0} + \nabla \times \Mb_{\textrm{gc},0}\right) = - \Bb_0 \times \frac{\nabla p_\textnormal{h}}{B_0} \times \bb_0 = - \nabla_\perp p_\textnormal{h}\,.
\end{equation}
This indicates that, with the spatially non-uniform distributions for which $\nabla_\perp p_\textnormal{h} \neq 0$, the linearized system \eqref{linear_system_1} includes a diamagnetic equilibrium current which might dominate the perturbed dynamics.

In order to mitigate this problem, rather than assuming the MHD equilibrium condition \eqref{equilibrium_1}, one should assume
$$
\Jb_0 \times \Bb_0 - \nabla p_0' = 0\,,\qquad p_0' =  p_0 + p_\textnormal{h} \,.
$$
That is to say, the pressure of the hot particles should be viewed as part of the MHD equilibrium pressure. However, we postpone a deeper discussion of this topic to future work. Here, in order to exclude the zeroth-order influence already at the linearization stage, we exclude the coupling terms with initial distribution $\fho$ but only consider the perturbed distribution $\tilde \fh = \fh - \fho$. Then the momentum balance equation of the second linearized scheme reads:
\begin{equation}\label{linear_system_2}
    \begin{aligned}
    \textnormal{Scheme 2 }\left\{
    \begin{aligned}
    \rho_0 \pdt{\wt \Ub} &- \Jb_0 \times \wt \Bb -  \frac{1}{\mu_0} (\nabla \times \wt \Bb) \times \Bb + \nabla \wt p 
    \\
    &- \left(\qh \ngc \wt \Ub -\Jgc(\tilde \fh,\Bb,\wt \Ub) - \nabla \times \Mgc(\tilde \fh) \right) \times \Bb = 0 \,, 
    \end{aligned}\right.
    \end{aligned}
\end{equation}
With this scheme, we define the following perturbed Hamiltonian where the particle energy with the initial distribution is excluded 
\begin{equation}\label{pHamiltonian2}
    \wt{\mathcal{H}}_2(t) = \int \frac{\rho_0}{2} |\wt \Ub|^2 \dx + \int \frac{1}{2\mu_0} |\wt \Bb|^2 \dx + \int \frac{\tilde p}{\gamma -1} \dx + \int \tilde \fh \frac{\mh}{2}\vp^2\dvp \dmu \dx + \int \tilde \fh \mu (\bb_0 \cdot \Bb)\dvp \dmu \dx  \,.
\end{equation}

In practice, Scheme 1 can be used when $\nabla p_\perp = 0$, as in spatially uniform distributions, whereas Scheme~2 becomes necessary when a density gradient is present (e.g., the numerical experiment in \autoref{sec5_itpa}). In the following, we mainly present the numerical scheme based on Scheme~1; analogous procedures also apply to Scheme~2, with the exception of the marker weights used when evaluating moments of the distribution, which will be detailed in \autoref{sec3_overview}.

\subsection{Normalization}\label{subsec2-3}
In what follows, let $\hat L$, $\hat B$ and $\hat n$ denote arbitrary units of length, magnetic field strength and number density. We normalize as
\begin{equation}
    \mathbf{x}=\mathbf{x}^\prime\hat{L}\,,\qquad \mathbf{B}=\mathbf{B}^\prime\hat{B}\,,\qquad \rho= A_\textnormal{b}m_\textnormal{p}n^\prime \hat{n}\,,
\end{equation}
where $A_\textnormal{b}m_\textnormal{p}$ is the mass of a bulk ion ($m_\textnormal{p}$ being the proton mass) and the primed quantities stand for unit-less numbers. Velocities are normalized as
\begin{equation}
    \mathbf{U}=\mathbf{U}^\prime\hat{v}_\textnormal{A}\,,\qquad \vp = \vp' \hat{v}_\textnormal{A}\,,\qquad \textrm{with} \qquad \hat{v}_\textnormal{A} = \frac{\hat B}{\sqrt{\mu_0 A_b m_p \hat{n}}}\,.
\end{equation}
For the remaining quantities, we set
\begin{equation}
t=t^\prime\hat \tau_\textnormal{A}\,,\quad \hat \tau_\textnormal{A} = \frac{\hat{L}}{\hat{v}_{\textnormal{A}}} \quad p=p^\prime\frac{\hat{B}^2}{\mu_0}\,,\quad  f_\textnormal{h}=f_\textnormal{h}^\prime\frac{\hat{n}}{\hat{v}_\textnormal{A} \hat{\mu}}\,,\quad \mu = \mu^\prime \hat{\mu}\,,\quad \hat \mu = \mu^\prime \frac{A_\textnormal{h} m_\textnormal{p} \hat{v}_\textnormal{A}^2}{\hat B} \,,
\end{equation}
where $A_\textnormal{h} m_\textnormal{p}$ stands for the mass of a hot particle.
Moreover, let us introduce the parameter $\varepsilon$ which relates the Alfvén time scale to the cyclotron frequency scale of the hot particles,
\begin{equation}
    \varepsilon = \frac{1}{\hat{\Omega}_\textnormal{ch}\hat \tau _\textnormal{A}}\,, \qquad \hat{\Omega}_\textnormal{ch} = \frac{Z_\textnormal{h}e\hat{B}}{A_\textnormal{h} m_\textnormal{p}}\,.
\end{equation}
With this, the normalized model equations read as follows, where the primes have been dropped and $A_\textnormal{h} / A_\textnormal{b}=~1$ is assumed for clarity:
\begin{subequations}\label{norm}
\begin{align}
        n_0 &\pdt{\wt \Ub} - (\nabla \times \Bb_0) \times \wt \Bb -  (\nabla \times \wt \Bb) \times \Bb + \nabla \wt p  -  \frac{1}{\varepsilon} \left(\ngc \wt \Ub - \Jgc - \varepsilon \nabla \times \Mgc \right) \times \Bb = 0 \,, \label{norm_hybrid_momentumbalance}
        \\[1mm]
        &\pdt{\tilde p}+ \nabla\cdot(p_0 \tilde{\mathbf{U}}) 
        + \frac{2}{3}\,p_0\nabla\cdot \tilde{\mathbf{U}}=0\,, \label{norm_hybrid_induction}
        \\
        &\pdt{\wt \Bb} + \nabla\times(\tilde{\mathbf{U}} \times \mathbf{B})
        = 0\,,\label{norm_hybrid_pressure}
        \\
        &\pdt \fh + \pder{}{\xb} \cdot \left[ \frac{\fh}{\Bpa} (\vp \Bb^* - \bb_0 \times \Eb^*)\right] + \frac{1}{\varepsilon} \pder{}{\vp} \left( \frac{\fh}{\Bpa} \Bb^* \cdot \Eb^*\right) = 0 \,.\label{norm_hybrid_driftkinetic}
\end{align}    
\end{subequations}
The normalized moments of the kinetic distribution are
\begin{subequations}
\begin{align}
    &\ngc(\fh) = \int \fh \dvp \dmu \,, 
    \\
    &\Mgc(\fh) = - \int \fh \mu \bb_0 \dvp \dmu \,, 
    \\
    &\Jgc(\fh,\Bb,\Ub) = \int \frac{\fh}{\Bpa} (\vp \Bb^* - \bb_0 \times \Eb^*) \dvp \dmu \,, \,
\end{align}
\end{subequations}
with the normalized effective electromagnetic fields 
\begin{equation}
    \Bb^* = \Bb + \varepsilon\,\vp\nabla \times \bb_0\,,\qquad \Eb^* = - \Ub \times \Bb - \varepsilon\,\mu\nabla (\bb_0 \cdot \Bb)\,.
\end{equation} 

\section{Spatial discretization}\label{sec3}
\subsection{Overview: FEEC-PIC discretization} \label{sec3_overview}
In this section, we derive a semi-discrete formulation of the model equations \eqref{norm}, i.e., a system of ODEs with continuous time but discrete space, that preserves the geometric and Hamiltonian structure. We adopt a FEEC-PIC framework: FEEC is used for the MHD equations \eqref{norm_hybrid_momentumbalance}-\eqref{norm_hybrid_pressure} and the PIC method is applied to the kinetic equation \eqref{norm_hybrid_driftkinetic} and the coupling terms. The overall strategy closely follows the structure-preserving discretization of the MHD-Vlasov system proposed in~\cite{Holderied_2021}, but is adapted here to driftkinetic particles, where the particle magnetic moment energy (last term in \eqref{Hamiltonian0}) couples the particles directly to the magnetic field. This coupling requires a particularly careful discretization so that later (in \autoref{sec3_Hamiltonian}) the discrete system can be assembled consistently into the skew-symmetric Hamiltonian structure.

\medskip \noindent \textbf{FEEC framework.}
All field variables are represented as proxy-functions of differential forms in curvilinear logical coordinates $\etab = (\eta_1, \eta_2, \eta_3) \in \hat \Omega$, where the reference domain is the unit cube $\hat \Omega = [0,1]^3$. The mapping to the physical domain $\xb = (x, y, z) \in \Omega$ is denoted by
\begin{equation}
    F:\quad \hat \Omega\, (\textnormal{logical}) \to \Omega\, (\textnormal{physical})\,,\quad (\eta_1, \eta_2, \eta_3) \mapsto  (x, y, z) \,,
\end{equation}
with Jacobian, metric tensor and metric determinant given by
\begin{equation}
    (DF)_{ij} = \pder{F_i}{\eta_j} \,, \quad G = DF^\top DF\,, \quad g = \text{det} \, G \,.
\end{equation}
\begin{figure}[t!p]
    \centering
    \includegraphics[width=1.\linewidth]{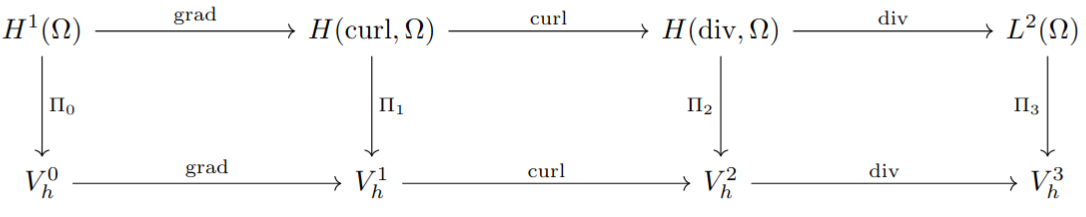}
    \caption{Commuting diagram of the de Rham complex. The top row depicts the continuous sequence of differential-form spaces forming an exact sequence, while the bottom row shows the corresponding finite-dimensional spaces. Projection operators commute with the differential operators, ensuring that the discrete sequence preserves the exactness of the continuous de Rham complex.}
    \label{fig:derham_diagram}
\end{figure}
The spaces of differential $k$-form proxy functions $\{H^1(\hat \Omega),\, H(\textnormal{curl, }\hat \Omega),\, H(\textnormal{div, }\hat \Omega),\, L^2(\hat \Omega) \}$ (defined in \ref{appendix_hilbert_spaces}) constitute the de Rham complex. The exact sequence of spaces ensures that the kernel of the next operator is the image of the previous one, i.e., Im(grad) = Ker(curl), Im(curl) = Ker(div) (the upper line in \autoref{fig:derham_diagram}). In this framework, each plasma variable is assigned to a differential form according to its geometric character:
\begin{itemize}
    \item kinetic distribution function $\fh$, MHD mass density $n$ and pressure $p$ are treated as volume-forms, as they represent volumetric densities;
    \item particle current $\Jgc$ and magnetic field $\Bb$ are 2-forms, since they are flux densities;
    \item particle Magnetization $\Mgc$ is a 1-form, since its curl acts as a current (2-form);
    \item MHD flow velocity $\Ub$ can, in principle, be represented as a 1-form, 2-form or a vector field in $H^1$; here we choose 2-form representation, which allows us to set essential boundary conditions for the normal component $\Ub \cdot \mathbf n$ to the boundary.
\end{itemize}
The pullback from the physical to the logical domain of 2-forms $\Bb \mapsto \hat \Bb^2$ and 3-forms $p \mapsto \hat p^3$ are given by
\begin{equation}
    \hat \Bb^2(\etab) = \sqrt g \,DF^{-1} \Bb(F(\etab))\,,\qquad \hat p^3(\etab) = \sqrt g\, p(F(\etab))\,.
\end{equation}

We are working with a conforming FE method, which means that the discrete spaces $V^0_h$, $V^1_h$, $V^2_h$, $V^3_h$ are subspaces of the continuous ones, as depicted in \autoref{fig:derham_diagram}. These discrete spaces will be constructed from tensor-product B-splines of high order; detailed definitions are given in \ref{appendix_discrete_spaces}. The discrete field variables are represented as
\begin{subequations}
    \begin{align}
        \hat \Ub^2 &\approx \hat \Ub_h^2(t,\etab) = \sum^3_{\mu=1}\sum_{ijk} u_{\mu,ijk} \Lambda^2_{\mu,ijk}(\etab) = \ub^\top \vec \Lambda^2 &&\in V^2_h \,,
        \\
        \hat \Bb^2 &\approx \hat \Bb_h^2(t,\etab) = \sum^3_{\mu=1}\sum_{ijk} b_{\mu,ijk} \Lambda^2_{\mu,ijk}(\etab)  = \bb^\top \vec \Lambda^2 &&\in V^2_h \,,
        \\[1mm]
        \hat p^3 &\approx \hat p_h^3(t,\etab) = \sum_{ijk} p_{ijk} \Lambda^3_{ijk} = \pb^\top \Lambda^3 &&\in V^3_h \,,
    \end{align}
\end{subequations}
where $\vec \Lambda^k := \textrm{diag}(\Lambda^k_{1,ijk}, \Lambda^k_{2,ijk}, \Lambda^k_{3,ijk}) \in \mathbb{R}^{N^2 \times 3}$ with $k=\{1,2\}$ collects the vector-valued basis functions, and lowercase bold letters denote vectors of finite-element coefficients of the corresponding variable, e.g., $\bb = (b_{\mu,ijk}) \in \mathbb{R}^{N^2}$. In addition, we introduce the discrete representation of gradient, curl and divergence, for instance,
\begin{equation}
    \hat \nabla \hat f^0 = (\vec \Lambda^1)^\top \mathbb{G} \mathbf{f}, \quad \hat \nabla \times \hat V^1 = (\vec \Lambda^2)^\top \mathbb{C} \mathbf{v}, \quad \hat \nabla \cdot \hat V^2 = \Lambda^3 \mathbb{D} \mathbf{v}\,,
\end{equation}
where $\mathbb{G}\in \mathbb{R}^{N^1 \times N^0}$, $\mathbb{C}\in \mathbb{R}^{N^2 \times N^1}$ and $\mathbb{D}\in \mathbb{R}^{N^3 \times N^2}$ satisfying $\mathbb{C} \mathbb{G} = 0$ and $\mathbb{D} \mathbb{C} = 0$. Another key ingredient of the de Rham diagram is the projection operators $\Pi_i,\, i=0,1,2,3$, which commute with the differential operators:
\begin{equation}
        \Pi_1\left[\hat \nabla \hat f^0\right]  = \hat \nabla \left(\Pi_0 \hat f^0\right) ,
        \quad
        \Pi_2\left[\hat \nabla \times \hat \Vb^1\right]  = \hat \nabla \times \left(\Pi_1 \hat \Vb^1\right) ,
        \quad
        \Pi_3\left[\hat \nabla \cdot \hat \Vb^2\right]  = \hat \nabla \left(\Pi_2 \hat \Vb^2\right)\,.
\end{equation}
Moreover, hat over the projectors $\hat \Pi_i$ refer to the coefficients obtained from a projection, 
\begin{equation}
    \Pi_2 \hat \Bb^2 = (\hat \Pi_2 \hat \Bb^2)^\top \vec \Lambda^2 = \bb^\top \vec \Lambda^2 = \hat \Bb^2_h\,.
\end{equation}

\medskip \noindent \textbf{PIC framework.}
The driftkinetic equation \eqref{norm_hybrid_driftkinetic} can be written in logical coordinates 
\begin{equation}\label{driftkinetic_logical}
\pdt \hfh + \pder{}{\etab} \cdot \left(\hfh \hat \ub_\textnormal{gc}\right) + \pder{}{\vp} \left(\hfh \hat a_\textnormal{gc}\right) = 0 \,,
\end{equation}
where the guiding-center velocity and acceleration are defined as
\begin{subequations}
\begin{align}
        \hat \ub_\textnormal{gc}(\etab, \vp) &=  \frac{1}{\hBpat (\etab, \vp)} \left[\vp \hBst(\etab, \vp) - \hbo(\etab) \times \hEso(\etab)\right] \,,
        \\
        \hat a_\textnormal{gc}(\etab, \vp) &= \frac{1}{\varepsilon} \frac{1}{\hBpat (\etab, \vp)} \hBst(\etab, \vp) \cdot \hEso(\etab) \,.
\end{align}
\end{subequations}
The kinetic distribution function, treated as a volume-form, is then approximated by using the Klimontovich representation ($N_p$ Lagrangian "markers" with delta-functions):
\begin{equation}\label{PIC_ansatz}
    \hfh (\etab,\vp,t) \approx \frac{1}{N_p} \sum^{N_p}_{p=1} \omega_p\, \delta(\etab - \etab_p(t)) \delta(\vp - v_{\parallel p}(t)) \,,
\end{equation}
and each marker evolves in time according to the equations of motion
\begin{subequations}\label{eom}
    \begin{align}
        \dt{\etab_p(t)} &= \hat \ub_\textnormal{gc}(\etab_p, v_{\parallel p}) &&\etab_p(t=0) = \etab_{p0}\,, \label{eom_eta}
        \\[1mm]
        \dt{v_{\parallel p}(t)} &= \hat a_\textnormal{gc}(\etab_p, v_{\parallel p}) &&v_{\parallel p}(t=0) = v_{\parallel p 0} \,. \label{eom_vp}
    \end{align}
\end{subequations}
In \eqref{PIC_ansatz}, we introduced the weight of each marker $\omega_p$, which, according to the Monte-Carlo sampling method, encodes the relation between the marker distribution $\hsh$ and the actual "physical" distribution function $\hfh$:
\begin{equation}
    \omega_p(t) := \frac{\hfh(\etab_p, v_{\parallel p}, t)}{\hsh(\etab_p, v_{\parallel p}, t)} \,.
\end{equation}
Here, the sampling density $\hsh$ is the volume-form probability density function (PDF) of markers such that
\begin{equation}
    \iiint \hsh (\etab, v_{\parallel}, t) \deta \dvp \dmu = 1 \qquad \forall t \in \mathbb{R}^+ \,.
\end{equation}
Assuming both $\hfh$ and $\hsh$ satisfy the same kinetic equation \eqref{driftkinetic_logical}, albeit with different initial conditions, $\omega_p$ remains constant along marker trajectories. As a result, integrals over distribution functions $\hfh$ can be approximated as Monte Carlo averages over markers, distributed according to the sampling density $\hsh$:
\begin{equation}\label{PIC_approximation}
    \begin{aligned}
        \iiint \hfh \hat Q \deta \dvp \dmu &= \iiint \frac{\hfh}{\hsh} \hat Q\, \hsh \deta \dvp \dmu \,,
        \\
        &\approx \frac{1}{N_p} \sum^{N_p}_{p=1} \omega_p \hat Q(\etab_p, v_{\parallel p}) \,.
        \end{aligned}
\end{equation}

The Monte-Carlo approach allows us to utilize several noise reduction techniques such as the $\delta f$-method. The main benefit of the $\delta f$-method comes from the reduction of statistical noise by replacing integrals over the known background distribution $f_0$ with analytical or numerical evaluations. In our scheme, this method is used to evaluate integrals involving the perturbed distribution function $\delta \hfh = \hfh - \hat f^\mr{vol}_{h0}$ in Scheme~2 \eqref{linear_system_2}:
\begin{equation}\label{PIC_approximation_deltaf}
    \begin{aligned}
        \iiint \delta \hfh \hat Q \deta \dvp \dmu &= \iiint \frac{\hfh - \hat f^\mr{vol}_{h0}}{\hsh} \hat Q\, \hsh \deta \dvp \dmu \,,
        \\
        &\approx \frac{1}{N_p} \sum^{N_p}_{p=1} \delta \omega_p \hat Q(\etab_p, v_{\parallel p}) \,,
    \end{aligned}
\end{equation}
where the "delta"-weight is defined as
\begin{equation}
    \begin{aligned}
        \delta \omega_p(t) &= \frac{\hfh(\etab_p, v_{\parallel p}, t) - \hat f^\mr{vol}_{h0}(\etab_p, v_{\parallel p}, t)}{\hsh(\etab_p, v_{\parallel p}, t)} \,,
        \\
        &= \omega_{p} - \frac{\hat f^\mr{vol}_{h0}(\etab_p, v_{\parallel p}, t)}{\hsh (\etab_p, v_{\parallel p}, t)} \,.
    \end{aligned}
\end{equation}

\subsection{Semi-discrete MHD equations} \label{sec3_MHD}
Prior to discretization, let us pull back the MHD equations to the logical domain. The momentum balance equation  \eqref{norm_hybrid_momentumbalance} is treated in weak formulation, while the induction \eqref{norm_hybrid_induction} and pressure \eqref{norm_hybrid_pressure} equations are kept in strong form. The weak formulation is obtained by taking the $L^2$-inner product with a test function $\hat \Cb^2 \in H(\textnormal{div, } \hat \Omega)$. After integration by parts, where boundary integrals are assumed to vanish, this yields
\begin{subequations}\label{me_logical}
    \begin{align}
        \left(\frac{\hat{n}^3_0}{\sg}\pdt{\hat \Ub^2},\, \hat \Cb^2\right)_2 
        = &\left(\hat \nabla \times (\frac{\hat \Bb_\mr{tot}^2}{\sg} \times \hat \Cb^2),\, \hat \Bb^2\right)_2 + \left(\Ginv \left[\hat \nabla \times (G \frac{\hat \Bb_0^2}{\sg})\times \hat \Bb^2 \right],\, \hat \Cb^2 \right)_2 + \left(\hat p^3, \hat \nabla \cdot \hat \Cb^2\right)_3 \nonumber
        \\
        &+ \underbrace{\left(\frac{1}{\varepsilon} \frac{\hngc}{\sg} \Ginv \left[\hat \Ub^2 \times \hat \Bb^2_\mr{tot}\right],\,  \hat \Cb^2\right)_2}_{\mathcal{C}(\hngc)} - \underbrace{\left(\frac{1}{\varepsilon} \Ginv \left[\hJgc \times \hat \Bb^2_\mr{tot}\right] ,\, \hat \Cb^2\right)_2}_{\mathcal{C}(\hJgc)} \label{me_logical_balance}
        \\
        &-\underbrace{\left(\hat \nabla \times (\frac{\hat \Bb^2_\mr{tot}}{\sg} \times \hat \Cb^2),\, \sg \Ginv \hMgc\right)_2}_{\mathcal{C}(\hMgc)} \,,  \nonumber
        \\
        \pdt{\hat \Bb^2} = &\hat \nabla \times (\hat \Ub^2 \times \frac{\hat \Bb_\mr{tot}^2}{\sg}) \,, \label{me_logical_induction}
        \\
        \pdt{\hat p^3} =  &- \hat \nabla \cdot(\frac{\hat p^3_0}{\sg} \hat \Ub^2) + (\gamma -1)\frac{\hat p^3_0}{\sg} \hat \nabla \cdot \hat \Ub^2 \,, \label{me_logical_pressure}
    \end{align}
\end{subequations}
where $\hat \Bb^2_\mr{tot} = \hat \Bb^2_0 + \hat \Bb^2$. The last three terms in \eqref{me_logical_balance}, $\mathcal{C}(\hngc)$, $\mathcal{C}(\hJgc)$ and $\mathcal{C}(\hMgc)$, denote three kinetic coupling terms involving the following particle moments:
\begin{subequations}
    \begin{align}
        \hngc &= \iint \hfh \dvp \dmu \,,
        \\
        \hJgc &= \iint \frac{\hfh}{\hBpat}(\vp \hat{\Bb}^{\ast2} - \hat \bb^1_0 \times \hat{\Eb}^{\ast1}) \dvp \dmu \,,
        \\
        \hMgc &= - \iint \frac{\hfh}{\sg} \mu \hat \bb^1_0 \,\dvp \dmu \,.
    \end{align}
\end{subequations}
Here, the 1-form pullback was used, $\hat \bb_0^1(\etab) = DF^\top \bb_0(F(\etab))$.

We now describe the semi-discretization of \eqref{me_logical} in space. First, we focus on the pure MHD part; the discrete formulation of the coupling terms will be presented in \autoref{sec3_driftkinetic}. By replacing the continuous variables with their discrete counterparts and applying the projection operators, we obtain the semi-discrete forms of \eqref{me_logical_induction} and \eqref{me_logical_pressure}:
\begin{subequations}
    \begin{align}
\dot{\bb}^\top \vec \Lambda^2 &= -\vec \Lambda^2 \mathbb{C} \hat \Pi^1 \left[ \frac{\hat{\Bb}^2_\mr{tot}}{\sqrt{g}} \times \vec \Lambda^2 \right]\ub \,, \\
\dot{\pb}^\top \Lambda^3 &= -\Lambda^3\mathbb{D} \hat \Pi^2\left[\frac{\hat p^3_0}{\sqrt{g}}\vec \Lambda^2\right] -(\gamma - 1) \Lambda^3\hat \Pi^3\left[\frac{\hat p^3_0}{\sqrt{g}} \Lambda^3\right]\mathbb{D} \ub \,.
    \end{align}
\end{subequations}
In the weak formulation, the $L^2$-inner product is represented as a weighted mass matrix, i.e.,
\begin{equation}
\begin{aligned}
    \left(\frac{\hat n^3_0}{\sg}\pdt{\hat \Ub^2},\, \hat \Cb^2\right)_2 
    &\approx \dot{\ub}^\top  \left(\int \frac{\hat n^3_0}{\sqrt g} \vec \Lambda^2 G \vec \Lambda^2 \frac{1}{\sqrt{g}} \, \deta \right) \cb \\
    &:= \dot{\mathbf{u}}^\top \mathbb{M}^{2,n} \cb \,.
\end{aligned}
\end{equation}
Altogether, the semi-discrete MHD system can be expressed in compact matrix-vector form as
\begin{subequations}\label{semi_discrete}
    \begin{align}
        \dot{\ub}^\top \mathbb{M}^{2,n}\cb &= \bb^\top\mathbb{M}^{2,J}\cb 
        + \bb^\top \mathbb{M}^{2}\mathbb{C}\mathcal{T}\cb 
        + \pb^\top\mathbb{M}^{3}\mathbb{D}\cb 
        + \mathcal{C}(\hngc) - \mathcal{C}(\hJgc) - \mathcal{C}(\hMgc)\,, \label{semi_discrete_balance} 
        \\[1mm]
        \dot \bb &= - \mathbb{C}\mathcal{T}\ub \,, \label{semi_discrete_induction} 
        \\[2mm]
        \dot \pb &= (-\mathbb{D}\mathcal{S} - (\gamma -1)\mathcal{K}\mathbb{D})\ub \,, \label{semi_discrete_pressure}
    \end{align}
\end{subequations}
with the following mass matrices:
\begin{subequations}
    \begin{align}
\mathbb{M}^2_{(\mu,ijk),(\nu,mno)} &:= \int \Lambda^2_{\mu,ijk} G \Lambda^2_{\nu,mno} \frac{1}{\sqrt{g}} \, \deta , \\
\mathbb{M}^{2,n}_{(\mu,ijk),(\nu,mno)} &:= \int \frac{\hat n^3_0}{\sqrt{g}} \Lambda^2_{\mu,ijk} G \Lambda^2_{\nu,mno} \frac{1}{\sqrt{g}} \, \deta , \\
\mathbb{M}^{2,J}_{(\mu,ijk),(\nu,mno)} &:= \int \Lambda^2_{\mu,ijk} (\hat \nabla \times \hat{\Bb}^2_0) \times \Lambda^2_{\nu,mno} \frac{1}{\sqrt{g}} \, \deta , \\
\mathbb{M}^3_{ijk,mno} &:= \int \Lambda^3_{ijk} \Lambda^3_{mno} \frac{1}{\sqrt{g}} \, \deta , \\
\end{align}
\end{subequations}
and the projection matrices: 
\begin{subequations}
    \begin{align}
\mathcal{T}_{(\mu,ijk),(\nu,mno)} &:= \hat \Pi^1_{{\mu,ijk}} \!\left[\frac{\hat{\Bb}^2_\mr{tot}}{\sqrt{g}} \times \vec \Lambda^2_{{\nu,mno}} \right] , \\
\mathcal{S}_{(\mu,ijk),(\nu,mno)} &:= \hat \Pi^2_{{\mu,ijk}} \!\left[\frac{\hat{p}^3_0}{\sqrt{g}} \vec \Lambda^2_{{\nu,mno}} \right] , \\
\mathcal{K}_{ijk,mno} &:= \hat \Pi^3_{ijk}\!\left[\frac{\hat{p}^3_0}{\sqrt{g}} \Lambda^3_{mno} \right] , \\
\mathcal{P}_{ijk,(\nu,mno)} &:= \hat{\Pi}^0_{ijk} \left[\frac{\hat \bb^1_0}{\sg} \cdot \vec \Lambda^2_{(\nu,mno)}\right].
\end{align}
\end{subequations}

\subsection{Coupling to driftkinetic particles} \label{sec3_driftkinetic}
We now turn to the semi-discrete formulation of the three coupling terms $\mathcal{C}(\hngc), \mathcal{C}(\hJgc)$ and $\mathcal{C}(\hMgc)$, appearing in \eqref{semi_discrete_balance}. The drift current contribution $\mathcal{C}(\hJgc)$ is first decomposed into three parts, each associated with a distinct guiding-center drift mechanism, namely the $E \times B$ drift current $\mathcal{C}(\hJgc)_{E\times B}$, the curvature drift current $\mathcal{C}(\hJgc)_{\nabla \times \bb}$ and the $\nabla B$ drift current $\mathcal{C}(\hJgc)_{\nabla B}$:
\begin{subequations}
\begin{align}
    \mathcal{C}(\hJgc)_{E\times B} &= -\frac{1}{\varepsilon} \iiint \frac{\hfh}{\sg} \left\{\frac{1}{\hBpat} \left[\hat \bb^1_0 \times \left( \hat \Ub^2 \times \frac{\hat \Bb^2_\mr{tot}}{\sg}\right)\right] \times \hat \Bb^2_\mr{tot} \right\} \cdot \frac{\hat \Cb^2}{\sg} \deta \dvp \dmu \,,
    \\
    \mathcal{C}(\hJgc)_{\nabla \times \bb} &= - \iiint \left(\frac{\hfh}{\hBpat} \vp^2 (\hat \nabla \times \hat \bb^1_0) \times \hat \Bb^2_\mr{tot} \right) \cdot \frac{\hat \Cb^2}{\sg} \deta \dvp \dmu \,,
    \\
    \mathcal{C}(\hJgc)_{\nabla B} &= - \iiint \frac{\hfh}{\hBpat} \left\{\hat \bb^1_0 \times \mu \hat \nabla \left( \hat \bb^1_0 \cdot \frac{\hat \Bb^2_\mr{tot}}{\sg} \right)\right\} \times \hat \Bb^2_\mr{tot} \cdot \frac{\hat \Cb^2}{\sg} \deta \dvp \dmu \,.
\end{align}
\end{subequations}
Then the density coupling $\mathcal{C}(\hngc)$ and the $E \times B$ drift current coupling $\mathcal{C}(\hJgc)_{E\times B}$ can be grouped together as
\begin{equation}\label{coupling_density}
    \mathcal{C}(\hngc) + \mathcal{C}(\hJgc)_{E \times B} =  \frac{1}{\varepsilon} \iiint \hfh \left(1 - \frac{\hat B_\parallel}{\hat B^*_\parallel}\right) \left( \hat \Ub^2 \times \frac{\hat \Bb^2_\mr{tot}}{\sg}\right) \cdot \frac{\hat \Cb^2}{\sg} \deta \dvp \dmu \,,
\end{equation}
where $\hat B_\parallel = (\hat \Bb^2_\mr{tot} \cdot \hat \bb^1_0)/\sqrt g$.
Note that these contributions are commonly disregarded in the literature, considering that two terms would cancel each other when one assumes $B^*_\parallel \approx B$. Upon applying the PIC approximation \eqref{PIC_approximation} with the discrete field variables, \eqref{coupling_density} is discretized as
\begin{equation}\label{coupling_density_discrete}
    \mathcal{C}(\hngc) + \mathcal{C}(\hJgc)_{E \times B} \approx 
    \frac{1}{\varepsilon}  \ub \frac{1}{N_p}\sum^{N_p}_{p=1} \omega_p \left(1 - \frac{\hat B_\parallel (\etab_p, v_{\parallel p})}{\hat B^{*}_\parallel (\etab_p, v_{\parallel p})}\right)\left( \vec \Lambda^2 (\etab_p) \times  \frac{\hat \Bb^2_\mr{tot}(\etab_p)}{\sg (\etab_p)} \right) \cdot\frac{\vec \Lambda^2  (\etab_p)}{\sg  (\etab_p)} \, \cb \,.
\end{equation}
Analogously, applying the same procedure yields the discrete forms for the remaining coupling terms: 
\begin{equation}\label{coupling_curvature_current_discrete}
        \mathcal{C}(\hJgc)_{\nabla \times b} \approx  -\frac{1}{N_p}\sum^{N_p}_{p=1} \frac{\omega_p}{\hBpat(\etab_p, v_{\parallel p})} v_{\parallel p}^2 (\hat \nabla \times \hat \bb^1_0(\etab_p)) \times \hat \Bb^2_\mr{tot}(\etab_p) \cdot \frac{\vec \Lambda^2 (\etab_p)}{\sqrt{g(\etab_p)}} \, \cb \,,
\end{equation}
\begin{equation}\label{coupling_gradB_current_discrete}
        \mathcal{C}(\hJgc)_{\nabla B} \approx - \frac{1}{N_p}\sum^{N_p}_{p=1} \frac{\omega_p}{\hBpat(\etab_p, v_{\parallel p})} \left\{ \hat \bb^1_0(\etab_p) \times \mu_p \vec \Lambda^1(\etab_p) \mathbb{G} \hat \Pi^0 \left[\hat \bb^1_0 \cdot \frac{\hat \Bb^2_\mr{tot}}{\sg}\right] \right\} \times \hat \Bb^2_\mr{tot}(\etab_p) \cdot \frac{\vec \Lambda^2 (\etab_p)}{\sqrt{g(\etab_p)}} \, \cb \,.
\end{equation}
and
\begin{equation}\label{coupling_magnetization_discrete}
    \begin{aligned}
        \mathcal{C}(\hMgc) \approx  \frac{1}{N_p} \sum^{N_p}_{p=1} \frac{\omega_p}{\sg} \mu_p \hat \bb_0^1(\etab_p) \vec \Lambda^2(\etab_p)  \mathbb{C} \mathcal{T} \cb \,.
    \end{aligned}
\end{equation}

With the coupling terms in place, the semi-discrete momentum balance equation \eqref{semi_discrete_balance} can be rewritten in compact matrix-vector form using the stacked notation introduced in \ref{appendix_stacked}:
\begin{equation}\label{semi_discrete_balace_mvform}
\begin{aligned}
        \dot{\ub}^\top \mathbb{M}^{2,n}\cb = &\bb^\top\mathbb{M}^{2,J}\cb + \bb^\top \mathbb{M}^{2}\mathbb{C}\mathcal{T}\cb + \pb^\top\mathbb{M}^{3}\mathbb{D}\cb -  \frac{1}{\varepsilon} \ub^\top \mathbb{L}^2 \bar W \left(\bar 1- \frac{\bar B_\parallel}{\bar B^{*}_\parallel}\right) \frac{1}{\bar g} \bar{\Bb}^\times_\mr{tot} \left( \mathbb{L}^2\right)^\top \cb \nonumber
        \\
        & +  \bar W \bar V_\parallel \frac{1}{\bar{B}^{*3}_\parallel}  \bar V_\parallel \bar{\Bb}^\times_\mr{tot} \overline{\nabla \times \bb_0}\frac{1}{\bar \sg} \left(\mathbb{L}^2\right)^\top \cb  +   \bar W  \frac{1}{\bar{B}^{*3}_\parallel} \bar{\Bb}^\times_\mr{tot} \bar{\bb}^\times_0  \bar M \overline{\nabla B_\parallel}_\mr{tot} \frac{1}{\bar \sg} \left( \mathbb{L}^2 \right)^\top \cb
        \\
        & +  \bar W \bar M \bar{\bb}_0 \frac{1}{\bar \sg}  \left(\mathbb{L}^2 \right)^\top \mathbb{C} \mathcal{T}  \cb \,.
\end{aligned}
\end{equation}
Finally, the equations of motion \eqref{eom} for all marker phase space coordinates $(\Hb, V_\parallel) \in \mathbb{R}^{3N_p + N_p}$ can also be expressed in a matrix-vector form:
\begin{subequations}\label{semi_discrete_eom}
    \begin{align}
        \dot \Hb &=\bar V_\parallel \left(\frac{1}{\bar{B}^{*3}_\parallel} \bar{\Bb}_\mr{tot} + \varepsilon \frac{1}{\bar{B}^{*3}_\parallel}  \bar V_\parallel \overline{\nabla \times \bb_0} \right) + \varepsilon \frac{1}{\bar{B}^{*3}_\parallel} \bar M \bar{\bb}^\times_0 \overline{\nabla B_\parallel}_\mr{tot} - \frac{1}{\bar{B}^{*3}_\parallel}  \bar{\bb}^\times_0 \bar{\Bb}^\times_\mr{tot} \left(\mathbb{L}^2\right)^\top \ub \,,
        \\
        \dot V_\parallel &=  - \left(\frac{1}{\bar{B}^{*3}_\parallel}\bar{\Bb}_\mr{tot} + \varepsilon \frac{1}{\bar{B}^{*3}_\parallel}\bar V_\parallel \overline{\nabla \times \bb_0} \right) \cdot \left( \bar M  \overline{\nabla B_\parallel}_\mr{tot}\right) - \left(\frac{1}{\bar{B}^{*3}_\parallel}\bar{\Bb}^\times_\mr{tot} \bar V_\parallel \overline{\nabla \times \bb_0} \right) \cdot \left(\frac{1}{\bar \sg} \left(\mathbb{L}^2\right)^\top \ub\right) \,.
    \end{align}
\end{subequations}

\subsection{Hamiltonian structure} \label{sec3_Hamiltonian}
To demonstrate that the semi-discrete system inherits the Hamiltonian structure of the continuous model, we introduce the discrete Hamiltonian functional. In terms of the discrete variables $\mathbf{Z}:=(\mathbf{u}, \mathbf{b}, \mathbf{p}, \Hb, V_\parallel) \in \mathbb{R}^{N^2 + N^2 + N^3 + 3N_p + N_p}$, the perturbed Hamiltonian \eqref{pHamiltonian1} becomes
\begin{equation}\label{discrete_energy}
\begin{aligned}
\tilde{\mathcal{H}}_h(\Zb) &= \frac{1}{2} \mathbf{u}^\top \mathbb{M}^{2,n} \mathbf{u} + \frac{1}{2} \mathbf{b}^\top \mathbb{M}^2\mathbf{b} + \frac{1}{\gamma - 1}\mathbf{p}^\top \mathbf{1}^3 + \frac{1}{2}V_\parallel W V_\parallel + M  W \bar{B}_{\parallel \mr{tot}}(\Hb, \bb) \,,
\end{aligned}
\end{equation}
where $\mathbf{1}^3 := (1, \cdots, 1) \in \mathbb{R}^{N^3}$ is a vector filled with ones.
The whole semi-discrete system \eqref{semi_discrete} and \eqref{semi_discrete_eom} can then be written as
\begin{equation}\label{semi_discrete_Hamiltonian}
\begin{aligned}
\frac{\textnormal{d} \mathbf{Z}}{\textnormal{d} t} &= \mathbb{J} \nabla_\mathbf{Z} \tilde{\mathcal{H}}_h + \mathbb{K} \Zb
\\
&= \overbrace{\begin{bmatrix}
\,\,\,\,\,J_{11} & J_{12} & 0 & \,\,\,\,\,J_{14} & J_{15} \\[1mm]
-J_{12}^\top & 0 & 0 & \,\,\,\,\,0 & 0 \\[1mm]
0 & 0 & 0 & \,\,\,\,\,0 & 0 \\[1mm]
-J_{14}^\top & 0 & 0 & \,\,\,\,\,J_{44} & J_{45} \\[1mm]
-J_{15}^\top & 0 & 0 & -J_{45}^\top & 0 \\[1mm]
\end{bmatrix}}^{:=\mathbb{J}}
\overbrace{\begin{bmatrix}
\mathbb{M}^{2,n}\mathbf{u} \\[1mm]
\mathbb{M}^2 \mathbf{b} + M  W (\mathbb{L}^0)^\top \mathcal{P}\\[1mm]
\frac{1}{\gamma - 1} \mathbf{1}^3 \\[1mm]
\bar M \bar W \overline{\nabla B}_{\parallel \mr{tot}} \\[1mm]
\bar W \bar V_\parallel
\end{bmatrix}}^{:=\nabla_\mathbf{Z} \tilde{\mathcal{H}}_h}
\\
&+ \underbrace{\begin{bmatrix}
0 & (\mathbb{M}^{2,n})^{-1}\mathbb{M}^{2,J} & (\mathbb{M}^{2,n})^{-1} \mathbb{D}^\top \mathbb{M}^3 & 0 & 0\\[1mm]
0 & 0 & 0 & 0 & 0\\[1mm]
- \left[ \mathbb{D} \mathcal{S} + (\gamma-1) \mathcal{K} \mathbb{D} \right] & 0 & 0 & 0 & 0 \\[1mm]
0 & 0 & 0 & 0 & 0\\[1mm]
0 & 0 & 0 & 0 & 0\\[1mm]
\end{bmatrix}}_{:=\mathbb{K}}
\underbrace{\begin{bmatrix}
    \ub \\[1mm] \bb \\[1mm] \pb \\[1mm] \Hb \\[1mm] V_\parallel
\end{bmatrix}}_{\Zb} \,,
\end{aligned}
\end{equation}
where the components of $\mathbb{J}$ are given by
\begin{equation}
\begin{aligned}
&J_{11}(\mathbf{b}, \Hb, V_\parallel) &&:= - \frac{1}{\varepsilon} \left(\mathbb{M}^{2,n}\right)^{-1} \mathbb{L}^2 \bar{W} \left(\bar 1- \frac{\bar B^0_\parallel}{\bar B^{*0}_\parallel}\right) \frac{1}{\bar g} {\bar{\Bb}}^\times_\mr{tot}  (\mathbb{L}^2)^\top \,,
\\
&J_{12}(\bb) &&:= (\mathbb{M}^{2,n})^{-1} \mathcal{T}^\top \mathbb{C}^\top \,,
\\
&J_{14}(\mathbf{b}, \Hb, V_\parallel) &&:= (\mathbb{M}^{2,n})^{-1} \mathbb{L}^2 \frac{1}{\bar \sg} \frac{1}{\bar{B}^{*3}_\parallel} \bar{\Bb}^\times_\mr{tot} \bar{\bb}^\times_0 \,,
\\
&J_{15}(\mathbf{b}, \Hb, V_\parallel) &&:= (\mathbb{M}^{2,n})^{-1} \mathbb{L}^2\frac{1}{\bar \sg} \frac{1}{\bar{B}^{*3}_\parallel}  \bar{\Bb}^\times_\mr{tot} \bar V_\parallel \overline{\nabla \times \bb_0} \,,
\\
&J_{44}(\mathbf{b}, \Hb, V_\parallel) &&:= \varepsilon \bar{W}^{-1} \frac{1}{\bar B^{*3}_\parallel}  \bar{\bb}^\times_0  \,,
\\
&J_{45}(\mathbf{b}, \Hb, V_\parallel) &&:=  \bar W^{-1}  \left(\frac{1}{\bar B^{*3}_\parallel} \bar{\Bb}_\mr{tot} + \frac{1}{\bar B^{*3}_\parallel} \varepsilon \bar{V}_\parallel \overline{\nabla \times \bb_0} \right) \,.
\\
\end{aligned}
\end{equation}
Here, the skew-symmetric so-called Poisson matrix $\mathbb{J}$ encodes a non-canonical Hamiltonian part, while $\mathbb{K}$ collects the non-Hamiltonian contributions. The skew-symmetric structure of $\mathbb{J}$ guarantees that the discrete Hamiltonian \eqref{discrete_energy} is conserved in time:
\begin{equation}
\frac{\textnormal{d}}{\textnormal{d} t} \tilde{\mathcal{H}}_h(\mathbf{Z}(t)) = (\nabla_\mathbf{Z} \tilde{\mathcal{H}}_h)^\top \frac{\textnormal{d} \mathbf{Z}}{\textnormal{d} t} = (\nabla_\mathbf{Z} \tilde{\mathcal{H}}_h)^\top \mathbb{J} \nabla_\mathbf{Z} \tilde{\mathcal{H}}_h = - (\nabla_\mathbf{Z} \tilde{\mathcal{H}}_h)^\top \mathbb{J} \nabla_\mathbf{Z} \tilde{\mathcal{H}}_h  = 0 \,.
\end{equation}
This property provides the foundation for designing energy-conserving time integrators, which will be the focus of the next section.

\section{Time discretization}\label{sec4}
\subsection{Poisson splitting and energy-conserving schemes}
To apply energy-preserving schemes, we perform Poisson splitting of the Hamiltonian part $\mathbb{J}\nabla_\Zb \tilde{\mathcal{H}}_h$. In this way, the skew-symmetric structure is preserved within each of the six sub-systems, here formally written as
\begin{equation}
    \mathbb{J} = \begin{bmatrix}
        J_{11}
    \end{bmatrix} + \begin{bmatrix}
        0 & J_{12} \\ - J_{12}^\top & 0
    \end{bmatrix} + \begin{bmatrix}
        0 & J_{14} \\ - J_{14}^\top & 0
    \end{bmatrix} + \begin{bmatrix}
        0 & J_{15} \\ - J_{15}^\top & 0
    \end{bmatrix} + \begin{bmatrix}
        J_{44}
    \end{bmatrix} + \begin{bmatrix}
        0 & J_{45} \\ - J_{45}^\top & 0
    \end{bmatrix} \,.
\end{equation}
We employ two types of energy-conserving time integrators, depending on the form of the discrete energy in each sub-step. In case of the sub-steps concerning the variables $\ub$, $\bb$ or $V_\parallel$, the implicit Crank-Nicolson method \cite{Crank_Nicolson_1947} is used, which exactly conserves linear or quadratic subsystem energies. For sub-steps involving the evolution of particle positions $\Hb$, we use the discrete gradient method \cite{McLachlan_1999} to preserve the discrete particle magnetic moment energy $M W \bar{B}_{\parallel \mr{tot}}(\Hb, \bb)$, which depends nonlinearly on $\Hb$. 

The discrete gradient method provides a general integral-preserving property, regardless of the form of the integral, for ODEs in skew-symmetric form, i.e.,
\begin{equation}
    \dot \zb = S(\zb) \nabla I(\zb) \quad \textnormal{with} \quad S(\zb)^\top = - S(\zb) \,.
\end{equation}
Then the system can be discretized as
\begin{equation}
    \frac{\zb^{n+1} - \zb^n}{\Delta t} = \bar S (\zb^{n}, \zb^{n+1}) \bar \nabla I(\zb^{n},\zb^{n+1}) \,,
\end{equation}
where $\bar S (\zb^{n}, \zb^{n+1})$ is any skew-symmetric matrix that converges to $S(\zb^n)$ when $\zb^{n+1} \rightarrow \zb^n$ and $\bar \nabla I(\zb^{n},\zb^{n+1})$ is a discrete gradient satisfying
\begin{align}
    (\zb^{n+1} - \zb^{n})\cdot \bar \nabla I(\zb^{n},\zb^{n+1}) &= I(\zb^{n+1}) - I(\zb^{n}) \,,
    \\
    \bar \nabla I(\zb^{n},\zb^{n}) &= \nabla I(\zb^{n}) \,.
\end{align}
Conservation of the integral $I$---in our case, the energy---follows directly:
\begin{equation}
    \begin{aligned}
        I(\zb^{n+1}) - I(\zb^{n}) &= (\zb^{n+1} - \zb^{n+1})^\top \bar \nabla I(\zb^{n},\zb^{n+1})
        \\
        &= \Delta t \bar \nabla I (\zb^{n},\zb^{n+1})^\top \bar S (\zb^{n},\zb^{n+1})^\top \bar \nabla I(\zb^{n},\zb^{n+1})
        \\
        &= -\Delta t \bar \nabla I (\zb^{n},\zb^{n+1})^\top \bar S (\zb^{n},\zb^{n+1}) \bar \nabla I(\zb^{n},\zb^{n+1})
        \\
        &= 0 \,.
    \end{aligned}
\end{equation}

\subsection{Implicit Crank-Nicolson method}\label{implicit_CN}
\medskip \noindent \textbf{Sub-step 1 (Density coupling).}
The first sub-step corresponds to the time evolution of $\wt \Ub$ due to the particle density $\ngc \wt \Ub \times \Bb$  and $E \times B$ part of the current coupling term $\Jgc \times \Bb$ in the momentum balance equation:
\begin{equation} \label{sub-system1}
    \dot \ub = J_{11} \mathbb{M}^{2,n} \ub \,.
\end{equation}
We solve \eqref{sub-system1} using the implicit Crank-Nicolson method:
\begin{equation}
    \frac{\ub^{n+1} - \ub^n}{\Delta t} = J_{11}\mathbb{M}^{2,n} \frac{\ub^{n+1} + \ub^n}{2} \,.
\end{equation}
Multiplying by $\mathbb{M}^{2,n}$ yields the linear system
\begin{equation}
    \left(\mathbb{M}^{2,n} - \frac{\Delta t}{2} \mathbb{M}^{2,n} J_{11} \mathbb{M}^{2,n} \right) \ub^{n+1} = \left( \mathbb{M}^{2,n} + \frac{\Delta t}{2} \mathbb{M}^{2,n} J_{11} \mathbb{M}^{2,n} \right) \ub^n \,.
\end{equation}

\medskip \noindent \textbf{Sub-step 2 (Shear Alfvén).}
The second sub-step describes Alfvénic dynamics, in which both $\wt \Ub$ and $\wt \Bb$ evolve. Guided by the Hamiltonian structure of the discrete system \eqref{semi_discrete_Hamiltonian}, the magnetization current contribution $(\nabla \times \Mgc) \times \Bb$ is naturally included, ensuring that the magnetization effect from particle gyro-motion is consistently coupled to the time evolution of Alfvén waves:
\begin{equation} \label{sub-system2}
    \begin{bmatrix}
        \dot \ub \\ \dot \bb
    \end{bmatrix}
    =
    \begin{bmatrix}
        0 & J_{12} \\ -J_{12}^\top & 0
    \end{bmatrix}
    \begin{bmatrix}
        \mathbb{M}^{2,n} \ub \\ \mathbb{M}^2 \mathbf{b} + M  W (\mathbb{L}^0)^\top \mathcal{P}
    \end{bmatrix} \,.
\end{equation}
The coupled system is advanced using the Crank-Nicolson method,
\begin{equation}\label{sub-system2_cn1}
    \begin{bmatrix}
        (\ub^{n+1} - \ub^n)/ \Delta t \\ (\bb^{n+1} - \bb^n)/\Delta t
    \end{bmatrix} =
    \begin{bmatrix}
        0 & J_{12}\\ - J_{12}^\top & 0
    \end{bmatrix}
    \begin{bmatrix}
        \mathbb{M}^{2,n} (\ub^{n+1} + \ub^n)/2 \\ \mathbb{M}^2 (\bb^{n+1} + \bb^n)/2 + M  W  (\mathbb{L}^0)^\top \mathcal{P}
    \end{bmatrix}\,,
\end{equation}
leading to the following linear systems
\begin{equation} \label{sub-system2_cn2}
    \begin{bmatrix}
        A & B \\ C & I
    \end{bmatrix} 
    \begin{bmatrix}
        \ub^{n+1} \\ \bb^{n+1} +   (\mathbb{M}^2)^{-1} M  W (\mathbb{L}^0)^\top \mathcal{P}
    \end{bmatrix} = 
    \begin{bmatrix}
        A & -B \\ -C & I
    \end{bmatrix} 
    \begin{bmatrix}
        \ub^{n} \\ \bb^{n} +   (\mathbb{M}^2)^{-1} M  W (\mathbb{L}^0)^\top \mathcal{P} \,,
    \end{bmatrix}
\end{equation}
where
\begin{equation}
    \begin{bmatrix}
        A & B \\ C & I
    \end{bmatrix} = 
    \begin{bmatrix}
        \mathbb{M}^{2,n} & -\Delta t/2 \mathbb{M}^{2,n}J_{12} \mathbb{M}^2 \\
        \Delta t/2 J_{12}^\top \mathbb{M}^{2,n} & I
    \end{bmatrix} \,.
\end{equation}
Using the Schur complement $S= A - B I^{-1}C$, the explicit update equations are then obtained:
\begin{subequations}
\begin{align}
    \ub^{n+1} &= S^{-1} \left[(A+BC) \ub^n -2B\left\{\bb^n +  (\mathbb{M}^2)^{-1} M  W  (\mathbb{L}^0)^\top \mathcal{P}\right\} \right] \,,
    \\
    \bb^{n+1} &= \bb^n - \frac{\Delta t}{2} J_{12}^\top \mathbb{M}^{2,n}(\ub^{n+1} + \ub^n) \,.
\end{align}
\end{subequations}

\medskip \noindent \textbf{Sub-step 3 (Current coupling $\nabla \times \bb$).}
The third sub-step involves direct energy exchange between the MHD and particle kinetic energies through the $\nabla \times \bb_0$ contribution in $\Jgc \times \Bb$ and the $\Bb^* \cdot \Eb^*$ parallel acceleration:
\begin{equation} \label{sub-system3}
    \begin{bmatrix}
        \dot \ub \\ \dot V_\parallel
    \end{bmatrix}
    =
    \begin{bmatrix}
        0 & J_{15} \\ -J_{15}^\top & 0
    \end{bmatrix}
    \begin{bmatrix}
        \mathbb{M}^{2,n} \ub \\ \bar W \bar V_\parallel
    \end{bmatrix} \,.
\end{equation}
We solve the system in the same way as sub-step 2
\begin{equation}
    \begin{bmatrix}
        (\ub^{n+1} - \ub^n)/ \Delta t \\ (V_\parallel^{n+1} - V_\parallel^n)/\Delta t
    \end{bmatrix} =
    \begin{bmatrix}
        0 & J_{15} \\ - J_{15}^\top & 0
    \end{bmatrix}
    \begin{bmatrix}
        \mathbb{M}^{2,n}(\ub^{n+1} + \ub^n)/2 \\  W (V_\parallel^{n+1} + V_\parallel^{n})/2
    \end{bmatrix}
\end{equation}
As in \eqref{sub-system2}, the Crank-Nicolson scheme is applied, leading to the update relations:
\begin{align}
    \ub^{n+1} &= S_3^{-1} \left[(\mathbb{M}^{2,n} - \frac{\Delta t^2}{4}\mathbb{M}^{2,n}J_{15}  W J_{15}^\top \mathbb{M}^{2,n})\ub^n + \Delta t \mathbb{M}^{2,n} J_{15}  W V_\parallel^n \right] \,,
    \\
    V^{n+1}_\parallel  &= V^n_\parallel - \frac{\Delta t}{2} J_{15}^\top \mathbb{M}^{2,n} (\ub^{n+1} + \ub^n) \,,
\end{align}
where $S_3 = \mathbb{M}^{2,n} + (\Delta t^2 / 4) \mathbb{M}^{2,n} J_{15}  W J_{15}^\top \mathbb{M}^{2,n}$.

\medskip \noindent \textbf{Sub-step 7 (Non-Hamiltonian part).}
The seventh sub-step concerns the time evolution of $\tilde \Ub$ and the MHD pressure $\tilde p$ associated with compressible waves and the first-order perturbation of the equilibrium Lorentz force $\Jb_0 \times \tilde \Bb$:
\begin{equation} \label{sub-system7}
    \begin{bmatrix}
        \dot \ub \\ \dot \pb
    \end{bmatrix}
    = \mathbb{K}
    \begin{bmatrix}
        \ub \\ \pb
    \end{bmatrix} \,.
\end{equation}
Although the sub-system does not have a skew-symmetric structure, we again apply the Crank-Nicolson method, leading to
\begin{equation}\label{sub-system7_cn1}
    \begin{bmatrix}
        (\ub^{n+1} - \ub^n)/ \Delta t \\ (\pb^{n+1} - \pb^n)/\Delta t
    \end{bmatrix} =
    \begin{bmatrix}
        0 & (\mathbb{M}^{2,n})^{-1} \mathbb{D}^\top \mathbb{M}^3 \\ -[\mathbb{D}\mathcal{S} + (\gamma - 1)\mathcal{K}\mathbb{D}] & 0
    \end{bmatrix}
    \begin{bmatrix}
        (\ub^{n+1} + \ub^n)/2 \\ (\pb^{n+1} + \pb^n)/2
    \end{bmatrix} + 
    \begin{bmatrix}
        (\mathbb{M}^{2,n})^{-1} \mathbb{M}^{2,J} \bb \\ 0
    \end{bmatrix} \,.
\end{equation}
Following the analogous procedures as the sub-step 2, the explicit update equations for $\ub^{n+1}$ and $\pb^{n+1}$ can be obtained as
\begin{equation}
\begin{aligned}
    \ub^{n+1} &= S_7^{-1} \left[\left\{\mathbb{M}^{2,n}-\frac{\Delta t^2}{4}\mathbb{D}^\top \mathbb{M}^3[\mathbb{D}\mathcal{S} + (\gamma - 1)\mathcal{K}\mathbb{D}]\right\} \ub^n + \Delta t \mathbb{D}^\top \mathbb{M}^3 \pb^n + \mathbb{M}^{2,J}\bb \right] \,,
    \\
    \pb^{n+1} &= \pb^n - \frac{\Delta t}{2}[\mathbb{D}\mathcal{S} + (\gamma - 1)\mathcal{K}\mathbb{D}](\ub^{n+1} + \ub^n) \,,
\end{aligned}
\end{equation}
where $S_7 = \mathbb{M}^{2,n} - (\Delta t^2/4)\mathbb{D}^\top \mathbb{M}^3[\mathbb{D}\mathcal{S} + (\gamma - 1)\mathcal{K}\mathbb{D}]$

\subsection{Discrete gradient method}\label{sec4:discrete_gradient}
In \texttt{STRUPHY}, the sub-steps 4--6 can be advanced either with explicit Runge-Kutta methods or with discrete gradient methods. While explicit methods are straightforward and flexible, they do not preserve energy. By contrast, the discrete gradient method introduced below provides exact conservation of non-quadratic energy. In what follows, we present the implementation details of the discrete gradient schemes for the individual sub-steps.

\medskip \noindent \textbf{Sub-step 5 (Driftkinetic $\boldsymbol{\nabla B}$).}
The fifth sub-step advances particle positions $\Hb$ due to the $\nabla B$ drift:
\begin{equation} \label{sub-system5}
        \dot \Hb = J_{44} \bar M \bar W \overline{\nabla B}_{\parallel \mr{tot}} \,.
\end{equation}
A second-order discrete gradient method is applied independently to each particle:
\begin{equation}\label{sub-system5_dg}
    \frac{\etab^{n+1}_p - \etab^{n}_p}{\Delta t} = \bar S(\etab^{n+1}_p, \etab^{n}_p) \bar \nabla I(\etab^{n+1}_p, \etab^{n}_p) \qquad p=1, \dots, N_p \,,
\end{equation}
with the skew-symmetric matrix
\begin{equation}
\begin{aligned}
    \bar S(\etab^{n+1}, \etab^{n}) &:= \varepsilon\frac{1}{\hBpat(\frac{\etab^{n+1} + \etab^{n}}{2})} \hat \bb^1_0(\frac{\etab^{n+1} + \etab^{n}}{2}) \times \,,
    \\
    &= \varepsilon\frac{1}{\hBpat(\frac{\etab^{n+1} + \etab^{n}}{2})} 
    \begin{bmatrix}
    0 & - \hat \bb^1_{0,3} & \hat \bb^1_{0,2} \\
    \hat \bb^1_{0,3} & 0 & - \hat \bb^1_{0,1} \\
    - \hat \bb^1_{0,2} & \hat \bb^1_{0,1} & 0
    \end{bmatrix}\,,
\end{aligned}
\end{equation} 
and a mid-point discrete gradient \cite{Gonzalez_1996} defined as
\begin{equation}
    \bar \nabla I(\etab^{n+1}, \etab^{n}) := \nabla I (\frac{\etab^{n+1} + \etab^{n}}{2}) + (\etab^{n+1} - \etab^{n})\frac{I(\etab^{n+1}) - I(\etab^{n}) - (\etab^{n+1} - \etab^{n})\cdot \nabla I(\frac{\etab^{n+1} + \etab^{n}}{2})}{||\etab^{n+1} - \etab^{n}||^2} \,.
\end{equation}
The scheme then preserves the integral $I$, denoting the magnetic moment energy of a single particle,
\begin{equation}
    I(\etab) = \mu \left[\hat B_0(\etab) + (\Lambda^0(\etab))^\top \mathcal{P} \bb \right] \in \mathbb{R} \,,
\end{equation}
where the gradient is given as
\begin{equation}
    \nabla I(\etab) = \mu \left[\hat \nabla \hat B_0 (\etab) +(\Lambda^1(\etab))^\top \mathbb{G} \mathcal{P} \bb \right] \in \mathbb{R}^3 \,.
\end{equation}
Consequently, it leads to conservation of the sub-system energy, which is simply the sum of all particle energies $M W \bar{B}_{\parallel \mr{tot}}(\Hb, \bb)$. The implicit update \eqref{sub-system5_dg} is solved by the standard fixed-point iteration:
\begin{equation}\label{sub-system5_fixed_point}
\begin{aligned}
    \etab^{n+1,0} &= \etab^{n} + \Delta t \bar S (\etab^n , \etab^n) \bar \nabla I(\etab^n, \etab^n) \,,
    \\
    \etab^{n+1,1} &= \etab^{n} + \Delta t \bar S (\etab^n , \etab^{n+1,0}) \bar \nabla I(\etab^n, \etab^{n+1,0}) \,,
    \\
    \dots
    \\
    \etab^{n+1,k} &= \etab^{n} + \Delta t \bar S (\etab^n , \etab^{n+1,k-1}) \bar \nabla I(\etab^n, \etab^{n+1,k-1}) \,,
\end{aligned}
\end{equation}
for $k = 0,1,\dots$ until satisfying
\begin{equation}
    ||\etab^{n+1,k} - \etab^{n+1,k-1}|| < \textnormal{tolerance} \,.
\end{equation}

\medskip \noindent \textbf{Sub-step 6 (Driftkinetic $\boldsymbol{B^*}$).}
The sixth sub-step concerns the evolution of both particle positions $\Hb$ and parallel velocities $V_\parallel$, coupled as
\begin{equation} \label{sub-system6}
    \begin{bmatrix}
        \dot \Hb \\ \dot V_\parallel
    \end{bmatrix}
    =
    \begin{bmatrix}
        J_{44} & J_{45} \\ -J_{45}^\top & 0
    \end{bmatrix}
    \begin{bmatrix}
        \bar M \bar W \overline{\nabla B}_{\parallel \mr{tot}} \\ \bar W \bar V_\parallel
    \end{bmatrix} \,.
\end{equation}
Unlike sub-step 5, convergence of the discrete gradient method is more challenging due to the coupled dynamics. We therefore employ the 1st order Itoh-Abe scheme \cite{Itoh_1988} with Newton-Raphson method:
\begin{equation} \label{sub-system6_dg}
    \frac{\zb^{n+1} - \zb^{n}}{\Delta t} = S(\zb^{n}) \bar \nabla I(\zb^{n+1}, \zb^{n}) \,,
\end{equation}
where $\zb := (\etab, v_\parallel) \in \mathbb{R}^{4}$ and the Itoh-Abe discrete gradient is defined as
\begin{equation}\label{sub-system6_dg_matrix}
    \bar \nabla I(\zb^{n+1}, \zb^{n}) := \begin{bmatrix}
        \frac{I(\eta_1^{n+1}, \eta_2^n, \eta_3^n, v_\parallel^n) - I(\eta_1^{n}, \eta_2^n, \eta_3^n, v_\parallel^n)}{\eta_1^{n+1} - \eta_1^n} \\
        \frac{I(\eta_1^{n+1}, \eta_2^{n+1}, \eta_3^n, v_\parallel^n) - I(\eta_1^{n+1}, \eta_2^n, \eta_3^n, v_\parallel^n)}{\eta_2^{n+1} - \eta_2^n} \\
        \frac{I(\eta_1^{n+1}, \eta_2^{n+1}, \eta_3^{n+1}, v_\parallel^n) - I(\eta_1^{n+1}, \eta_2^{n+1}, \eta_3^n, v_\parallel^n)}{\eta_3^{n+1} - \eta_3^n} \\
        \frac{I(\eta_1^{n+1}, \eta_2^{n+1}, \eta_3^{n+1}, v_\parallel^{n+1}) - I(\eta_1^{n+1}, \eta_2^{n+1}, \eta_3^{n+1}, v_\parallel^n)}{v_\parallel^{n+1} - v_\parallel^n}
    \end{bmatrix} \,,
\end{equation}
with the integral
\begin{equation}
    I(\zb) = \mu \left[\hat B_0(\etab) + (\Lambda^0(\etab))^\top \mathcal{P} \bb \right] + \frac{1}{2} v_{\parallel}^2 \in \mathbb{R}\,.
\end{equation}
Then the equation \eqref{sub-system6_dg} can be considered as a following root $(F=0)$ finding problem
\begin{equation}
    F(\zb^{n+1}) := \zb^{n+1} - \zb^{n} - \Delta t S(\zb^{n}) \bar \nabla I(\zb^{n+1}, \zb^{n})\,.
\end{equation}
From the initial guess
\begin{equation}
    \zb^{n+1,0} = \zb^{n} + \Delta t S(\zb^{n}) \nabla I(\zb^{n}) \,,
\end{equation}
we iteratively solve the equation
\begin{equation}
    \zb^{n+1,k+1} = \zb^{n+1,k} - J_F^{-1}(\zb^{n+1,k}) F(\zb^{n+1,k}) \,,
\end{equation}
where the Jacobian of F is given as
\begin{equation}
    [J_F(\zb^{n+1,k})]_{i,j} = \frac{\partial F_i(\zb^{n+1,k})}{\partial \zb^{n+1,k}_j} = \begin{pmatrix}
        1 & 0 & 0 & 0 \\
        0 & 1 & 0 & 0 \\
        0 & 0 & 1 & 0 \\
        0 & 0 & 0 & 1
    \end{pmatrix} - \Delta t S(\zb^n) \frac{\partial \bar \nabla I_i(\zb^{n+1,k}, \zb^n)}{\partial \zb^{n+1,k}_j} \,,
\end{equation}
where $i,j = 0,1,2,3$.
Note that the Jacobian of the discrete gradient \eqref{sub-system6_dg_matrix} does not include second order derivatives since the Itoh-Abe discrete gradient \eqref{sub-system6_dg_matrix} is a derivative-free scheme. Therefore, the Jacobian of the discrete gradient can be readily obtained
\begin{equation}
    \begin{aligned}
        \frac{\partial \bar \nabla I_i(\zb^{n+1,k}, \zb^n)}{\partial \eta^{n+1,k}_1} &= \begin{bmatrix}
            \frac{\nabla I_1(\eta^{n+1,k}_1, \eta^n_2, \eta^n_3, v^n_\parallel)}{\eta^{n+1,k}_1 - \eta^n_1} - \frac{I(\eta^{n+1,k}_1, \eta^n_2, \eta^n_3, v^n_\parallel) - I(\eta^{n}_1, \eta^n_2, \eta^n_3, v^n_\parallel)}{(\eta^{n+1,k}_1 - \eta^n_1)^2} \\
            \frac{\nabla I_2(\eta^{n+1,k}_1, \eta^{n+1,k}_2, \eta^n_3, v^n_\parallel) - \nabla I_2(\eta^{n+1,k}_1, \eta^n_2, \eta^n_3, v^n_\parallel)}{\eta^{n+1,k}_2 - \eta^n_2} \\
            \frac{\nabla I_3(\eta^{n+1,k}_1, \eta^{n+1,k}_2, \eta^{n+1,k}_3, v^n_\parallel) - \nabla I_3(\eta^{n+1,k}_1, \eta^{n+1,k}_2, \eta^n_3, v^n_\parallel)}{\eta^{n+1,k}_3 - \eta^n_3} \\
            0
        \end{bmatrix} \,,
        \\
                \frac{\partial \bar \nabla I_i(\zb^{n+1,k}, \zb^n)}{\partial \eta^{n+1,k}_2} &= \begin{bmatrix}
            0 \\
            \frac{\nabla I_2(\eta^{n+1,k}_1, \eta^{n+1,k}_2, \eta^n_3, v^n_\parallel)}{\eta^{n+1,k}_1 - \eta^n_1} - \frac{I(\eta^{n+1,k}_1, \eta^{n+1,k}_2, \eta^n_3, v^n_\parallel) - I(\eta^{n+1,k}_1, \eta^n_2, \eta^n_3, v^n_\parallel)}{(\eta^{n+1,k}_1 - \eta^n_1)^2} \\
            \frac{\nabla I_3(\eta^{n+1,k}_1, \eta^{n+1,k}_2, \eta^{n+1,k}_3, v^n_\parallel) - \nabla I_3(\eta^{n+1,k}_1, \eta^{n+1,k}_2, \eta^n_3, v^n_\parallel)}{\eta^{n+1,k}_3 - \eta^n_3} \\
            0
        \end{bmatrix} \,,
        \\
                \frac{\partial \bar \nabla I_i(\zb^{n+1,k}, \zb^n)}{\partial \eta^{n+1,k}_3} &= \begin{bmatrix}
            0 \\
            0 \\
            \frac{\nabla I_3(\eta^{n+1,k}_1, \eta^{n+1,k}_2, \eta^{n+1,k}_3, v^n_\parallel)}{\eta^{n+1,k}_3 - \eta^n_3} - \frac{I(\eta^{n+1,k}_1, \eta^{n+1,k}_2, \eta^{n+1,k}_3, v^n_\parallel) - I(\eta^{n+1,k}_1, \eta^{n+1,k}_2, \eta^n_3, v^n_\parallel)}{(\eta^{n+1,k}_3 - \eta^n_3)^2} \\
            0
        \end{bmatrix} \,,
        \\
                \frac{\partial \bar \nabla I_i(\zb^{n+1,k}, \zb^n)}{\partial 
                v_\parallel^{n+1,k}} &= \begin{bmatrix}
            0 \\
            0 \\
            0 \\
            \frac{1}{2}
        \end{bmatrix} \,,
    \end{aligned}
\end{equation}
where
\begin{equation}
    \nabla I (\zb) = 
    \begin{bmatrix}
        \mu \left[\hat \nabla \hat B_0(\etab) + (\Lambda^1(\etab))^\top \mathbb{G} \mathcal{P} \bb\right] \\
        v_{\parallel}
    \end{bmatrix}\in \mathbb{R}^4 \,.
\end{equation}

\medskip \noindent \textbf{Sub-step 4 (Current coupling $\nabla B$).}
The fourth sub-step couples $\wt \Ub$ to the particle positions $\Hb$:
\begin{equation} \label{sub-system4}
    \begin{bmatrix}
        \dot \ub \\ \dot \Hb
    \end{bmatrix}
    =
    \begin{bmatrix}
        0 & J_{14} \\ -J_{14}^\top & 0
    \end{bmatrix}
    \begin{bmatrix}
        \mathbb{M}^{2,n} \ub \\ \bar M \bar W \overline{\nabla B}_{\parallel \mr{tot}}
    \end{bmatrix} \,.
\end{equation}
We again apply the 2nd order mid-point discrete gradient method as the scheme used for \eqref{sub-system5_dg}:
\begin{equation}\label{sub-system4_dg}
    \frac{\zb^{n+1} - \zb^{n}}{\Delta t} = \bar S(\zb^{n+1}, \zb^{n}) \bar \nabla I(\zb^{n+1}, \zb^{n}) \,,
\end{equation}
where $\Zb:= (\ub, \Hb) \in \mathbb{R}^{N^2 + 3N_p}$ and skew-symmetric matrix is given as
\begin{equation}
    \bar S(\Zb^{n+1}, \Zb^{n}) = 
    \begin{bmatrix}
        0 & J_{14}(\bb^{n}, \frac{\Hb^{n+1} + \Hb^{n}}{2}, V_\parallel^{n})
        \\
        - J_{14}^\top(\bb^{n}, \frac{\Hb^{n+1} + \Hb^{n}}{2}, V_\parallel^{n}) & 0
    \end{bmatrix} \,.
\end{equation}
The discrete gradient is defined as
\begin{equation}
    \bar \nabla I(\Zb^{n+1}, \Zb^{n}) = \nabla I (\frac{\Zb^{n+1} + \Zb^{n}}{2}) + (\Zb^{n+1} - \Zb^{n})\frac{I(\Zb^{n+1}) - I(\Zb^{n}) - (\Zb^{n+1} - \Zb^{n})\cdot \nabla I(\frac{\Zb^{n+1} + \Zb^{n}}{2})}{||\Zb^{n+1} - \Zb^{n}||^2} \,,
\end{equation}
with the subsystem energy integral,
\begin{equation}
    I(\Zb) = \frac{1}{2}\ub^\top \mathbb{M}^{2,n} \ub +  M W  \bar{B}_{\parallel \mr{tot}}(\Hb)  \,,
\end{equation}
and its gradient,
\begin{equation}
    \nabla I(\Zb) = \begin{bmatrix}
        \mathbb{M}^{2,n} \ub \\[1mm]
         M W \overline{\nabla B_\parallel}_\mr{tot} (\Hb)
    \end{bmatrix}\in \mathbb{R}^{N^2 + 3N_p} \,.
\end{equation}

Unlike two driftkinetic steps \eqref{sub-system5} and \eqref{sub-system6}, where each particle converges to its own independent solution, all particles are mutually correlated through $\Ub$. In other words, we seek a solution for the set of unknowns~$\Zb$, consisting of the finite-element coefficients $\ub$ and positions of all PIC markers $\Hb$. For this reason, convergence of the standard fixed-point iteration is not always guaranteed; hence, the relaxed fixed-point iteration \cite{Ehrhardt_2024} is used, with a relaxation factor $\theta \in [0,1]$:
\begin{equation}\label{sub-system4_fixed_point}
\begin{aligned}
    \Zb^{n+1,0} &= \Zb^{n} + \Delta t \bar S (\Zb^n , \Zb^n) \bar \nabla I(\Zb^n, \Zb^n) \,,
    \\
    \Zb^{n+1,1} &= (1 - \theta)\Zb^{n+1,0} + \theta\left[\Zb^{n} + \Delta t \bar S (\Zb^n , \Zb^{n+1,0}) \bar \nabla I(\Zb^n, \Zb^{n+1,0})\right] \,,
    \\
    \dots
    \\
    \Zb^{n+1,k} &=(1 - \theta)\Zb^{n+1,k-1} + \theta\left[\Zb^{n} + \Delta t \bar S (\Zb^n , \Zb^{n+1,k-1}) \bar \nabla I(\Zb^n, \Zb^{n+1,k-1})\right] \,,
\end{aligned}
\end{equation}
for $k = 0,1,\dots$ until satisfying
\begin{equation}
    ||\Zb^{n+1,k} - \Zb^{n+1,k-1}|| < \textnormal{tolerance} \,.
\end{equation}
While this method provides exact conservation of non-quadratic energy, the convergence of the scheme requires that the integral and its gradient remain smooth throughout the computational domain, including boundaries. In toroidal geometry, a bounded radial coordinate violates this condition. Therefore, in the toroidal experiments of \autoref{sec5_itpa}, sub-step 4 \eqref{sub-system4} is advanced using the standard explicit integrator.

\section{Numerical experiments}\label{sec5}
\subsection{Verification of energy conservation}\label{sec5_energy_conservation}
\begin{figure}[t!p]
    \centering
    \input{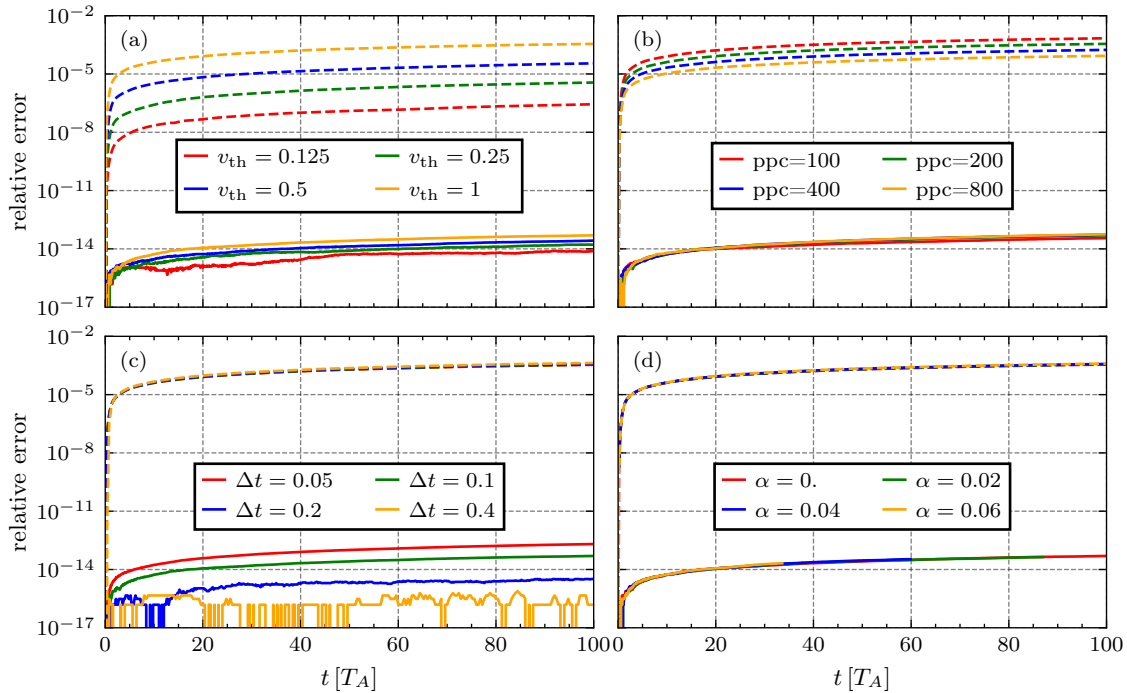}
    \caption{Time evolution of the relative error in total energy for different simulation parameters: (a) kinetic thermal velocity~\texorpdfstring{$v_{\mathrm{th}}$}{v\_th}, (b) number of particles per cell (ppc), (c) time step size \texorpdfstring{$\Delta t$}{Delta t} and (d) mesh distortion factor \texorpdfstring{$\alpha$}{alpha}. Results from the discrete gradient scheme (solid lines) are compared with the fourth-order Runge-Kutta scheme (dashed lines). }
    \label{fig:ch5-1_fig1}
\end{figure}
In \autoref{sec4}, we introduced two energy-preserving time integrators: the implicit Crank-Nicolson scheme for linear or quadratic energy and the discrete gradient scheme for non-quadratic energy. The conservation property of the Crank–Nicolson scheme in hybrid MHD–kinetic systems has already been demonstrated in \cite{Holderied_2021}. In this section, we extend the analysis to non-quadratic energies and assess the conservation properties of the discrete gradient scheme.

In order to have nontrivial driftkinetic dynamics, spatial non-uniformity of the magnetic field is required. We therefore consider a sheared periodic equilibrium magnetic field,
\begin{equation}
    \Bb(\xb) = B_0 \left(\eb_z + \frac{L_x}{q(\xb)} \eb_y\right)\,, \qquad q(\xb) = q_0 + q_1\sin\left(2\pi \frac{\xb}{L_x}\right) \,,
\end{equation}
with the Colella mesh distortion mapping for slab geometry,
\begin{equation}
    \Fb_\textnormal{Colella}: \hat \Omega \rightarrow \Omega, \quad \etab = \begin{bmatrix}
        \eta_1 \\ \eta_2 \\ \eta_3
    \end{bmatrix} \mapsto
    \xb = \begin{bmatrix}
        L_x [ \eta_1 + \alpha \sin(2\pi\eta_1)\sin(2\pi\eta_2)] \\ 
        L_y [ \eta_2 + \alpha \sin(2\pi\eta_2)\sin(2\pi\eta_3)] \\ 
        L_z \eta_3
    \end{bmatrix} \,,
\end{equation}
where $(L_x, L_y, L_z)$ are the side lengths of the periodic box and $0 \leq \alpha \leq 1/(2\pi)$ is a mesh-distortion factor introducing geometric and metric effects. The periodicity of the magnetic field is chosen to ensure compatibility with the discrete gradient method by guaranteeing smoothness of the preserved energy. Kinetic particles are initialized with a spatially uniform Maxwellian distribution. To isolate exact energy conservation, the non-Hamiltonian step \eqref{sub-system7} is switched off. For the fixed-point iteration in the discrete gradient scheme, we set the tolerance to $10^{-13}$ and use a relaxation factor $\theta = 0.5$ in sub-step 4 \eqref{sub-system4}.

Simulations are performed in a periodic slab of length $L_x=20$, $L_y=L_z=40\pi$, with the equilibrium parameters $B_0=1\,T$, $q_0=1$ and $q_1=0.5$. The spatial discretization employs $N_\textnormal{el} = (24,24,24)$ spline elements with degree $p=(3,3,3)$. \autoref{fig:ch5-1_fig1} shows the relative error in total energy from the discrete gradient scheme compared to the explicit fourth-order Runge-Kutta method under systematic parameter scans. Starting from a default setup ($v_\textnormal{th}=v_A$, ppc = 200, $\Delta t = 0.1\,T_A$ and $\alpha = 0$), each subplot varies a single parameter. \autoref{fig:ch5-1_fig1}(a) demonstrates that the discrete gradient scheme conserves total energy within the solver tolerance $10^{-13}$ for all thermal velocities, while the explicit scheme exhibits energy errors increasing with $v_\textnormal{th}$. \autoref{fig:ch5-1_fig1}(b) shows that errors in the explicit scheme increase with lower ppc, whereas the discrete gradient scheme maintains energy conservation regardless of ppc. \autoref{fig:ch5-1_fig1}(c)-(d) further confirms that its conservation property is unaffected by variations in $\Delta t$ or $\alpha$. These results confirm that the discrete gradient scheme preserves the exact energy balance during nonlinear particle–field interactions and is unaffected under physical and numerical parameter variations.

\subsection{wave-particle interactions in toroidal geometry}\label{sec5_itpa}
\subsubsection{Mapping and boundary conditions}
As a second numerical test, we examine the wave-particle interactions in toroidal geometry. Exploiting toroidal symmetry, the simulation domain covers only $1/n_\text{tor}$ of hollow torus, mapped as
\begin{equation}
    \Fb_\textnormal{Torus}: \hat \Omega \rightarrow \Omega, \quad \etab = \begin{bmatrix}
        \eta_1 \\ \eta_2 \\ \eta_3
    \end{bmatrix} \rightarrow
    \xb = \begin{bmatrix}
        [\{a_1 + (a_2-a_1)\eta_1\}\cos(2\pi\eta_2)+ R]\cos(2\pi\eta_3/n_\text{tor}) \\ 
        [\{a_1 + (a_2-a_1)\eta_1\}\cos(2\pi\eta_2)+ R]\sin(-2\pi\eta_3/n_\text{tor}) \\ 
        \{a_1 + (a_2-a_1)\eta_1\}\sin(2\pi\eta_2)
    \end{bmatrix} \,,
\end{equation}
where $a_1$ is the possible inner-hole radius around the magnetic axis and $a_2$ and $R$ are minor and major radii of the torus. To handle the boundary of the torus, clamped B-splines with Dirichlet boundary conditions are used for the radial coordinate $\eta_1$. In contrast, periodic splines are used for $\eta_2$ and $\eta_3$ representing poloidal and toroidal angles. For the kinetic part, special care is required for the time evolution of PIC marker positions. In sub-step 4 \eqref{sub-system4} and 5 \eqref{sub-system5}, if PIC markers touch the boundary during iteration, the iteration is terminated and the markers are either removed from the simulation or refilled at the opposite poloidal angle $\theta_\text{refill} = -\theta_\text{lost}$ and corresponding toroidal angle of the same magnetic flux surface $\phi_\text{refill} = -2q(r_\text{lost})\theta_\text{lost}$ where $r_\text{lost}$ and $\theta_\text{lost}$ denote radial position and poloidal angle of the lost particles. In sub-step 6 \eqref{sub-system6}, as noted at the end of \autoref{sec4:discrete_gradient}, the classical fourth-order Runge-Kutta method is employed.

\subsubsection{Numerical stability and filtering}
Self-consistent hybrid MHD–kinetic codes are known as being prone to numerical instabilities, since the PIC noise is inherent to the particle discretization and nonlinear field–particle interactions can further amplify these fluctuations. In many existing codes, such instabilities are mitigated by introducing artificial dissipation and applying Fourier filtering to suppress unwanted modes. 

Unconditional stability is observed in our scheme when the non-Hamiltonian step \eqref{sub-system7} is disabled. When this step is included, the scheme generally remains stable without artificial dissipation, but numerical instabilities may arise from the non-Hamiltonian part; these instabilities originate from the linearization of MHD, leading to an artificial injection of surplus energy. In the present test, we turn on the non-Hamiltonian step to include compressive plasma responses and equilibrium pressure effects. However, the first-order perturbation of the equilibrium Lorentz force, $\Jb_0 \times \tilde{\Bb}$, is disabled to minimize an accumulation of surplus energy. With this measure, simulations remain stable without artificial dissipation.

In addition, two types of filters are applied to the coupling terms \eqref{coupling_density_discrete}-\eqref{coupling_magnetization_discrete}. A binomial three-point smoothing filter is always applied to suppress grid-scale noise such as finite grid instabilities, which are inherent to FEM–PIC coupling. For instance, filtering is applied to the coefficients of the accumulated (from particles) 0-form field 
\begin{equation}
    \hat A = \frac{1}{N_p}\sum^{N_p}_{p=1} \omega_p Q_p \Lambda^0(\etab_p) \in V^0_\textnormal{h}
\end{equation}
yielding
\begin{equation}
    \hat A^{0,\textnormal{filtered}}_{i,j,k} = \sum^2_{l_1=0}\sum^2_{l_2=0}\sum^2_{l_3=0} S(l_1) S(l_2) S(l_3) \hat A^{0}_{i-1+l_1,j-1+l_2,k-1+l_3} \,,
\end{equation}
where $S = \frac{1}{4}[1,2,1]$ is the one-dimensional binomial mask. As an additional option, a toroidal Fourier filter is also implemented, 
\begin{equation}
    \hat A^{0,\textnormal{filtered}} = \mathcal{F}^{-1} [ \mathcal{F}[\hat A](k)]\,,
\end{equation}
where $\mathcal{F}$ is the one-dimensional discrete Fourier transformation along the toroidal direction $\eta_3$. This filter retains only a prescribed toroidal mode ($n=k$) of the kinetic contributions, suppressing all others, including the zeroth-order component. Unlike the binomial filter, the Fourier filter is not introduced for noise control but only for comparison. As will be demonstrated in \autoref{sec5_itpa_results}, comparison between filtered and unfiltered cases allows us to assess the impact of non-target spectral components on the observed wave–particle interactions.

\subsubsection{ITPA benchmark case}\label{sec5_itpa_results}
To verify the scheme in toroidal geometry, we adopt the ITPA benchmark case \cite{Könies_2018}, which is designed to study linear interactions between TAEs and EPs. The setup considers two plasma species in a circular tokamak of large aspect ratio (the major radius $R=10\,m$, the minor radius $a=1\,m$). The bulk species is a hydrogen plasma with a flat density profile $n = 2.0 \cdot 10^{19} m^{-3}$ and constant temperature T=1 keV, while the equilibrium pressure decreases towards the boundary: $p_0(r) = 7.17 \cdot 10^{13} - 6.811 \cdot 10^{3}r^2 - 3.585 \cdot 10^{2} r^4$ Pa. The corresponding ad-hoc MHD equilibrium is constructed with circular concentric flux surfaces given by $\Bb_0 = \nabla \psi \times \nabla \phi + F\nabla \phi $ where the poloidal flux $\psi$ satisfies
\begin{equation}\label{poloidal_flux}
    \dt \psi = \frac{B_\text{axis} r}{q(r)\sqrt{1 - (\frac{r}{R})^2}} \,,
\end{equation}
with safety factor profile $q(r) = 1.71 + 0.16 (r/a)^2$, poloidal current function $F=-B_\text{axis} R$ and on-axis magnetic field $B_\text{axis} = 3\,T$. The second species is energetic deuterons, described by a Maxwellian distribution with radial density profile
\begin{equation}\label{ITPA_EPdensity_profile}
    n_\textrm{h}(r) = n_0 c_3 \exp \left(- \frac{c_2}{c_1} \tanh{\frac{r-c_0}{c_2}}\right) \,,
\end{equation}
where $n_0 = 1.44131 \cdot 10^{17} m^{-3}$ and the profile coefficients are $c_0 = 0.49123$, $c_1 = 0.49123$, $c_2 = 0.198739$ and $c_3 = 0.521298$. However, a local Maxwellian distribution, which is a local thermodynamic equilibrium of particles, is not the true equilibrium in the toroidal geometry, and relaxation to the true equilibrium would reduce the density gradient. To avoid this, we adopt the canonical Maxwellian distribution, following the approach used in \cite{Lu_2023, Angelino_2006}. The distribution is then given by
\begin{equation}
    \fh(\psi_{0\textnormal{shift}}, \epsilon, \mu) = \frac{n_\textnormal{h}(\psi_{0\textnormal{shift}})}{(2\pi)^{3/2} v_\text{th}^3} \exp \left(-\frac{\epsilon}{v^2_\text{th}}\right) \,,
\end{equation}
which depends on the constants of motion of particles in an axisymmetric equilibrium magnetic field: energy~$\epsilon$, magnetic moment~$\mu$ and shifted toroidal angular momentum, 
\begin{equation}
    \psi_{0\textnormal{shift}} = \psi + \frac{\mh F}{\qh B_0}v_\parallel - \textnormal{sign}(v_\parallel)\frac{\mh}{\qh}R\sqrt{2(\epsilon - \mu B_\textnormal{axis})} H(\epsilon - \mu B_\textnormal{axis}) \,,
\end{equation}
where $H$ is the Heaviside function.

We performed a parameter scan of the EP temperature $T_\textnormal{h}$ over the range 100--800 keV. For each $T_\textnormal{h}$, two simulations were carried out; one with the $n=6$ toroidal Fourier filter and one without. 
\begin{figure}[t!p]
    \centering
    \resizebox{0.9\linewidth}{!}{\input{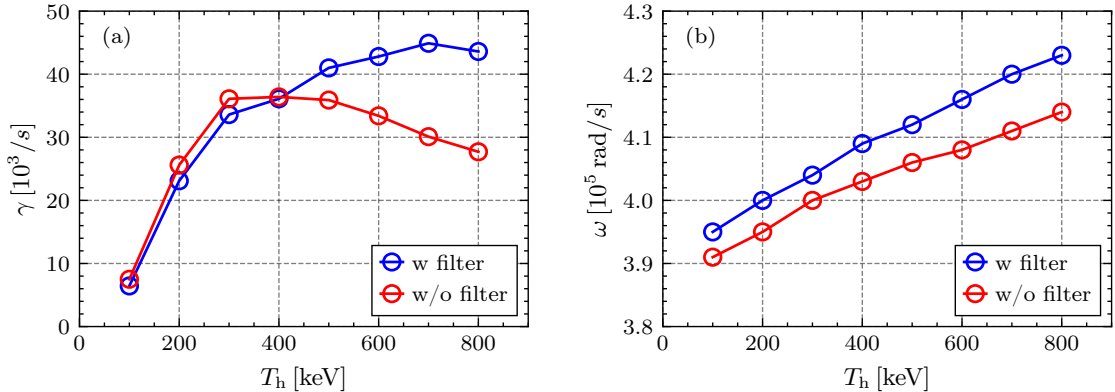}}
    \caption{
    Growth rates of TAEs (left) and corresponding mode frequencies (right) as functions of EP temperature, obtained from simulations with (blue) and without (red) toroidal Fourier filter.
    }
    \label{fig:ch5-2_fig1}
\end{figure}
\autoref{fig:ch5-2_fig1}(a) shows the resulting TAE growth rates. The filtered results show a good agreement with other codes reported in \cite{Könies_2018}, whereas clear deviations appear in the unfiltered case for $T_\textnormal{h} > 400$ keV. The corresponding TAE mode frequencies are plotted in \autoref{fig:ch5-2_fig1}(b).  The filtered cases are in close agreement with the MEGA code (see Figure 2 of \cite{Könies_2018}), and the unfiltered results yield slightly lower frequencies. 
\begin{figure}[t!p]
    \centering
    \resizebox{0.8\linewidth}{!}{\input{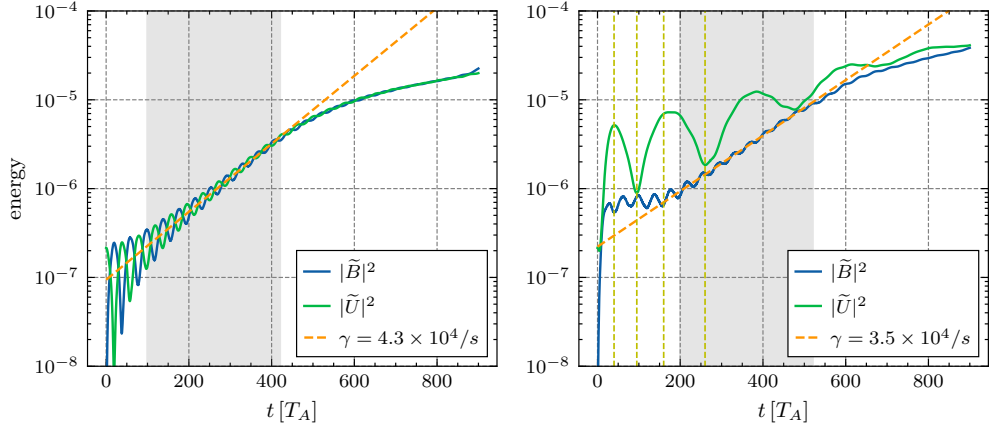}}
    \caption{
    The time evolution of magnetic (blue) and MHD kinetic (green) energies with (left) and without (right) toroidal Fourier filter. The growth rate is estimated from linear fits (orange dashed lines) during the linear growth phase (shaded gray areas).
    }
    \label{fig:ch5-2_fig2}
\end{figure}
The time evolution of magnetic and MHD kinetic energies for $T_\textnormal{h}$ = 500 keV is presented in \autoref{fig:ch5-2_fig2}.  In the filtered case (left), a distinct linear growth phase begins around $t \sim 100\,T_A$. In contrast, the unfiltered case (right) exhibits finite-amplitude fluctuations in the MHD kinetic energy from the very beginning of the simulation. These initial perturbations obscure the linear TAE growth, which can only be identified after the TAE energy overtakes the initial perturbations at $t \sim 200\,T_A$. In addition, the unfiltered case exhibits a reduced growth rate, as also presented in \autoref{fig:ch5-2_fig2}(a).

\begin{figure}[t!p]
    \centering
    \resizebox{0.8\linewidth}{!}{\input{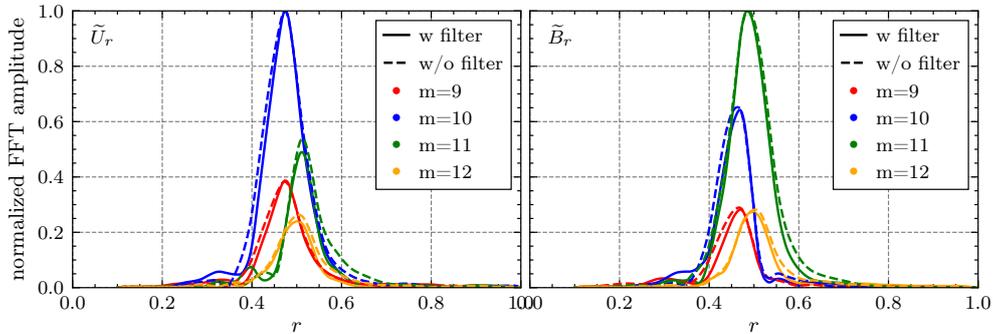}}
    \caption{
    The radial mode structures of the perturbed radial MHD velocity (left) and magnetic field (right) for toroidal mode $n=6$ during the linear growth phase $(t=300\,T_A)$. The four dominant poloidal harmonics are plotted $(m=9,10,11,12)$. Solid and dashed lines correspond to the results with and without the toroidal Fourier filter, respectively.
    }
    \label{fig:ch5-2_fig3}
\end{figure}
\autoref{fig:ch5-2_fig3} shows the radial mode structures of the perturbed radial MHD velocity and magnetic field at the linear growth phase ($t=300\,T_A$). For the MHD velocity, the $m=10$ harmonic is dominant followed by $m=11$, whereas for the magnetic field $m=11$ dominates with $m=10$ as the second largest component. The radial location and mode width are comparable to results from other codes, despite the latter presenting electrostatic potential $\phi$ rather than electromagnetic variables. The present simulations also exhibit more pronounced minor harmonics ($m=9, \,12$), a tendency observed in codes employing ad-hoc MHD equilibrium. No significant differences are found between the filtered and unfiltered cases.
\begin{figure}[t!p]
    \centering
    \includegraphics[width=.9\linewidth]{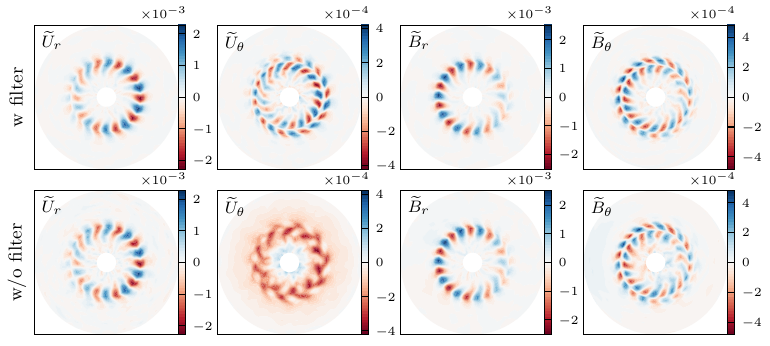}    
    \caption{The poloidal mode structures of radial and poloidal components of the perturbed MHD velocity and magnetic field during the linear growth phase $(t=300\,T_A)$. Upper row: with toroidal Fourier filter; lower row: without filter.
    }
    \label{fig:ch5-2_fig4}
\end{figure}
The corresponding 2D mode structures are presented in \autoref{fig:ch5-2_fig4}. In the filtered case (upper row), the radial components ($\widetilde U_r, \, \widetilde B_r$) exhibit a clear mode structure with characteristic poloidal asymmetry, and the poloidal components ($\widetilde U_\theta, \, \widetilde B_\theta$) display a spiral-like pattern with radially flipped phases, which reflects the polarization of the wave and the incompressibility constraint of the dynamics. The overall structures are identical in both filtered and unfiltered runs (lower row), except for the poloidal MHD velocity, where a radially localized zonal flow ($m=0$ shear flow) is superimposed on the TAEs. This flow corresponds to the MHD kinetic energy fluctuations in \autoref{fig:ch5-2_fig2} and exhibits GAM-like (geodesic acoustic mode) oscillations with the frequency $\omega \sim 0.2\,\omega_\textnormal{TAE}$. 
\begin{figure}[t!p]
    \centering
    \includegraphics[width=.9\linewidth]{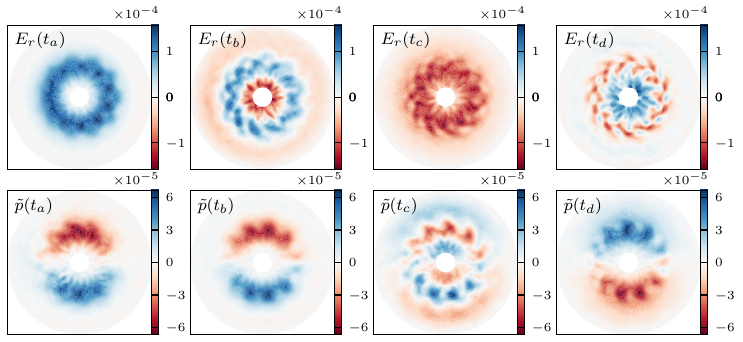}    
    \caption{
    The poloidal mode structures of radial electric field and perturbed pressure at the four times ($t_a=40\,T_A$, $t_b=95\,T_A$, $t_c=160\,T_A$ and $t_d=260\,T_A$) without toroidal Fourier filter. The selected times correspond to successive extrema of the MHD kinetic energy (yellow dashed lines in the right plot of \autoref{fig:ch5-2_fig2}).
    }
    \label{fig:ch5-2_fig5}
\end{figure}
Four oscillation phases are illustrated in the contour plots of the radial electric field and the perturbed pressure (\autoref{fig:ch5-2_fig5}). The upper row shows the oscillatory $m=0$ structure, and the characteristic $m=1$ up-down pressure mode is observed in the lower row. Further studies are needed to clarify the origin of such zonal flows and their impact on TAE dynamics.
\begin{figure}[t!p]
    \centering
    \includegraphics[width=.9\linewidth]{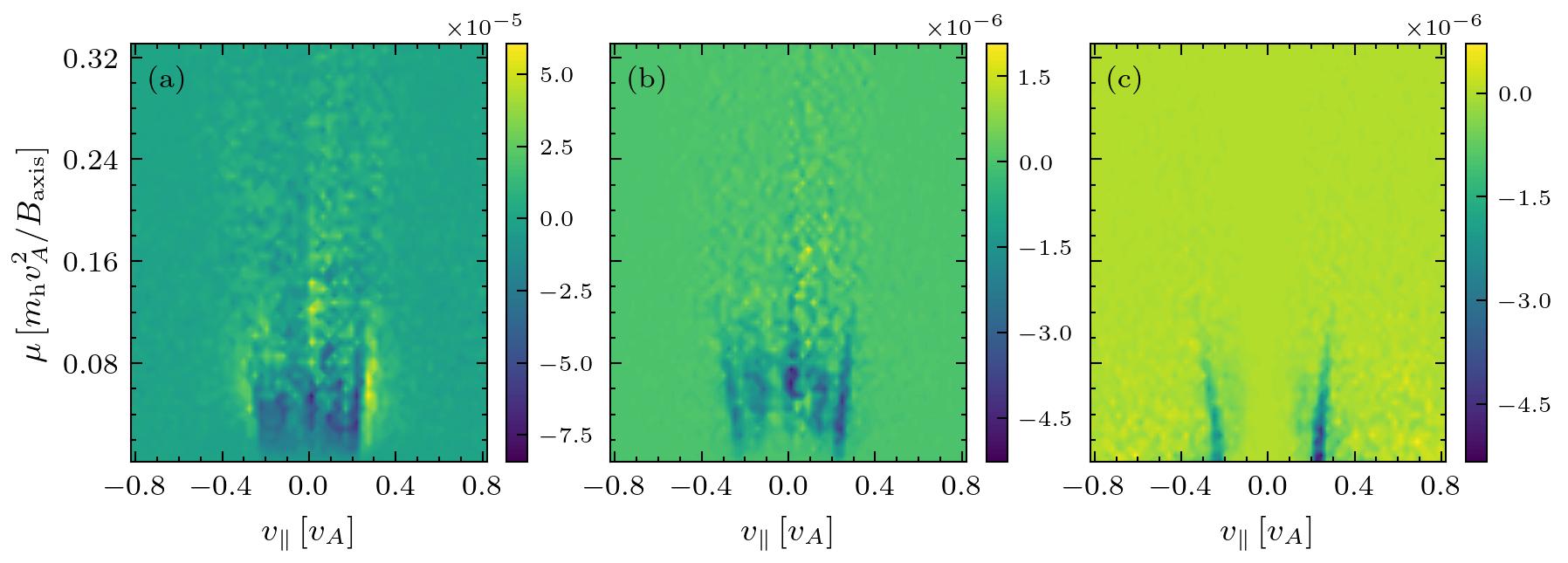}    
    \caption{
    Accumulated sum of spatially averaged particle energy differences, $\Delta E_\textnormal{p} = E_\textnormal{p}(t + \Delta t) - E_\textnormal{p}(t)$, during the linear growth phase, binned in parallel velocity and magnetic moment. Three plots correspond to energy transfer at three different coupling sub-steps of the time integrator: (a) sub-step 2 (magnetization current), (b) sub-step 4 ($\nabla B$ current) and (c) sub-step~3 ($\nabla \times \bb$ current).
    }
    \label{fig:ch5-2_fig6}
\end{figure}
We additionally examine the net energy exchange between EPs and the wave by accumulating particle energy difference binned in parallel velocity and magnetic moment, as shown in \autoref{fig:ch5-2_fig6}. \autoref{fig:ch5-2_fig6}(a) corresponds to sub-step 2 \eqref{sub-system2}, where magnetization current of particles is included in the time evolution of Alfvénic dynamics, and no clear resonance is observed. In contrast, \autoref{fig:ch5-2_fig6}(b) and (c), which represent the two drift current contributions, display apparent resonant interactions at the specific parallel velocities.

\section{Conclusions and outlook}\label{sec6}
This work extends the structure-preserving discretization of the MHD–Vlasov system proposed in \cite{Holderied_2021} to the driftkinetic regime, where the larger time step allows efficient simulations of low-frequency wave-particle interactions. The model equations derived from a variational principle yield a skew-symmetric discrete Poisson matrix, where the energy-conserving schemes can be naturally applied. A key novelty is the inclusion of particle magnetic moment energy arising from the guiding-center approximation, a non-quadratic energy term depending simultaneously on particle positions and magnetic field. The tailored discretization of the term derives the sub-systems: one consistently couples the guiding-center magnetization to the evolution of Alfvén dynamics, and another relates particle drifts induced by the perturbed electric field to the MHD flow driven by the kinetic drift current. On top of this, we demonstrated how a discrete gradient scheme can be adapted to conserve the non-quadratic term, employing the Itoh-Abe scheme with the Newton-Raphson method and relaxed fixed-point iterations. We showed that the resulting methods conserve energy exactly in periodic domains regardless of numerical or physical parameters, including metric distortions. The scheme further enables simulations of EP-driven TAEs growth without artificial dissipation or mode filtering. It thus may provide a pathway toward high-fidelity simulations in regimes where artificial dissipation would otherwise degrade small-scale dynamics or where the spectral mode cannot be prescribed in advance, such as nonlinear multi-mode dynamics. Moreover, \texttt{STRUPHY} code is openly available, and the parameter files used for the present simulations are provided in~\cite{STRUPHY_simulations}, allowing reproducibility of the results.

Several directions for future work emerge from this study. From a modeling perspective, the systematic procedure adopted here---a model derived from a variational principle followed by FEEC–PIC discretization---opens the possibility of reliable comparison between MHD–Vlasov and MHD–driftkinetic models. Since both are treated within the same modeling and numerical framework, differences can be attributed to physics alone, thereby allowing us to quantify the impact of the guiding-center approximation on energetic particle dynamics. In this context, since the present work employs the current-coupling scheme, a further step is to extend the formulation to pressure-coupling models, such as those proposed in \cite{Tronci_2010, Close_2018}, enabling systematic comparison between the two coupling approaches. Another direction is to move beyond the perturbative setting---linearized MHD on the prescribed equilibrium---and adopt a fully nonlinear, non-perturbative hybrid formulation in which equilibrium and fluctuations evolve self-consistently, including the equilibrium pressure and current contributions of the kinetic species; in realistic tokamak geometries, an understanding of such hybrid equilibria remains open. Yet, such an extension may even be favorable for an energy-preserving scheme, since it would remove the non-Hamiltonian part introduced by linearization. Finally, developing discrete gradient methods applicable to bounded integrals in non-periodic geometry may represent a promising research direction, as they would allow exact energy conservation in tokamak simulations.

\appendix
\section{Hilbert spaces and $L^2$-inner product}\label{appendix_hilbert_spaces}
The four Hilbert spaces for differential k-forms are defined as
\begin{equation}
\begin{alignedat}{2}
    H^1(\hat \Omega) &:= \{\hat f^0: \hat \Omega \rightarrow \mathbb{R}, \quad &&\textnormal{s.t.}\quad (\hat f^0,\hat f^0)_0 < \infty\,, \quad (\hat \nabla \hat f^0, \hat \nabla \hat f^0)_1 < \infty\}\,,
    \\
    H(\textnormal{curl, }\hat \Omega) &:= \{\hat \Vb^1: \hat \Omega \rightarrow \mathbb{R}^3, \quad &&\textnormal{s.t.}\quad (\hat \Vb^1,\hat \Vb^1)_1 < \infty\,, \quad (\hat \nabla \times \hat \Vb^1, \hat \nabla \times \hat \Vb^1)_2 < \infty\}\,,
    \\
    H(\textnormal{div, }\hat \Omega) &:= \{\hat \Vb^2: \hat \Omega \rightarrow \mathbb{R}^3, \quad &&\textnormal{s.t.}\quad (\hat \Vb^2,\hat \Vb^2)_2 < \infty\,,\quad (\hat \nabla \cdot \hat \Vb^2, \hat \nabla \cdot \hat \Vb^2)_3 < \infty\}\,,
    \\
    L^2(\hat \Omega) &:= \{\hat f^3: \hat \Omega \rightarrow \mathbb{R}, \quad &&\textnormal{s.t.} \quad (\hat f^3,\hat f^3)_3 < \infty \}\,,
\end{alignedat}
\end{equation}
where the $L^2$-inner product is given by:
\begin{equation}
    (\hat a^p, \hat b^p)_p := 
    \left\{ 
    \begin{aligned}
        &\int_{\hat \Omega} \hat a^0 \hat b^0 \sg \deta\,, && p=0\,,
        \\
        &\int_{\hat \Omega} (\hat a^1)^\top G^{-1} \hat b^1 \sg \deta\,, && p=1\,,
        \\
        &\int_{\hat \Omega} (\hat a^2)^\top G \hat b^2 \oosg \deta\,, && p=2\,,
        \\
        &\int_{\hat \Omega} \hat a^3 \hat b^3 \oosg \deta\,, && p=3\,.
    \end{aligned}
    \right.
\end{equation}
\section{Three-dimensional tensor products of B-splines}\label{appendix_discrete_spaces}
Four discrete spaces are spanned by tensor products of the B-spline basis functions $N^p_i$ and their differentials $D^{p-1}_i$:
\begin{subequations}
    \begin{align}
        V^0_h &= \textnormal{span}\left\{\left.\Lambda^0_{ijk}(\etab) := N^p_i(\eta_1)N^p_j(\eta_2) N^p_k(\eta_3)\right. \left.\right| 0 \leq i < n^1_N, 0 \leq j < n^2_N,  0 \leq k < n^3_N \right\} \,,
\\
\textnormal{grad} \, & \, \bigg\downarrow\nonumber\\
        V^1_h &= \textnormal{span}
        \left\{ \left.
        \begin{aligned}
        \Lambda^1_{1,ijk}(\etab):= D^{p-1}_i(\eta_1)N^p_j(\eta_2) N^p_k(\eta_3) 
        \\
        \Lambda^1_{2,ijk}(\etab):= N^p_i(\eta_1)D^{p-1}_j(\eta_2) N^p_k(\eta_3)
        \\
        \Lambda^1_{3,ijk}(\etab) := N^p_i(\eta_1)N^p_j(\eta_2) D^{p-1}_k(\eta_3) 
        \end{aligned}
        \right. \right| \left.
        \begin{aligned}
             0 \leq i < n^1_D, 0 \leq j < n^2_N,  0 \leq k < n^3_N,
            \\
            0 \leq i < n^1_N, 0 \leq j < n^2_D,  0 \leq k < n^3_N,
            \\
            0 \leq i < n^1_N, 0 \leq j < n^2_N,  0 \leq k < n^3_D
        \end{aligned}\right\} \,,
\\
\textnormal{curl} \, & \, \bigg\downarrow\nonumber\\
        V^2_h &= \textnormal{span}
        \left\{\left.
        \begin{aligned}
        \Lambda^2_{1,ijk}(\etab):= N^p_i(\eta_1)D^{p-1}_j(\eta_2) D^{p-1}_k(\eta_3)
        \\
        \Lambda^2_{2,ijk}(\etab) := D^{p-1}_i(\eta_1)N^p_j(\eta_2) D^{p-1}_k(\eta_3)
        \\
        \Lambda^2_{3,ijk}(\etab) := D^{p-1}_i(\eta_1)D^{p-1}_j(\eta_2) N^p_k(\eta_3)
        \end{aligned}
        \right. \right| \left.
        \begin{aligned}
            0 \leq i < n^1_N, 0 \leq j < n^2_D,  0 \leq k < n^3_D,
            \\
            0 \leq i < n^1_D, 0 \leq j < n^2_N,  0 \leq k < n^3_D,
            \\
            0 \leq i < n^1_D, 0 \leq j < n^2_D,  0 \leq k < n^3_N
        \end{aligned}\right\} \,,
\\
\textnormal{div} \, & \, \bigg\downarrow\nonumber\\
        V^3_h &= \textnormal{span}\left\{\left.
        \Lambda^3_{ijk}(\etab) :=D^{p-1}_{i_1}(\eta_1)D^{p-1}_{i_2}(\eta_2) D^{p-1}_{i_3}(\eta_3) \right| 0 \leq i < n^1_D, 0 \leq j < n^2_D,  0 \leq k < n^3_D\right\}\,,
    \end{align}
\end{subequations}
where $n^{\mu=1,2,3}_{N/D}$ denotes the number of splines in each direction. In the main text, $N^n$ with $n\in\{0,1,2,3\}$ represents the total number of basis functions in each space, and $N^n_\mu$ for $n\in\{1,2\}$ and $\mu\in\{1,2,3\}$ denotes the number of basis functions for each component of the vector-valued spaces, e.g., $N^2 = N^2_1 + N^2_2 + N^2_3 = n^1_N n^2_D n^3_D + n^1_D n^2_N n^3_D + n^1_D n^2_D n^3_N$.

\section{Stacked vector and matrix}\label{appendix_stacked}
The discretized coupling terms, including particle moments \eqref{coupling_density_discrete}-\eqref{coupling_magnetization_discrete} and the equations of motion for all markers \eqref{semi_discrete_eom}, can be compactly written with the following stacked (along the number of markers $N_p$) vector and matrix notations:
\begingroup
\allowdisplaybreaks
\begin{alignat*}{3}
& \boldsymbol{\Hb} &&:= (\eta_{1,1}, \dots, \eta_{1,N_p},\eta_{2,1}, \dots, \eta_{2,N_p},\eta_{3,1}, \dots, \eta_{3,N_p}) \qquad &&\in \mathbb{R}^{3N_p} \,,
\\
& V_\parallel &&:= (v_{\parallel,1}, \dots v_{\parallel,N_p}) &&\in \mathbb{R}^{N_p} \,,
\\
&\bar V_\parallel &&:= \mr{diag}(V_\parallel) \otimes I_{3\times3} &&\in \mathbb{R}^{3N_p \times 3N_p} \,,
\\
&M &&:= (\mu_{1}, \dots \mu_{N_p}) &&\in \mathbb{R}^{N_p} \,,
\\
&\bar M &&:= \mr{diag}(M) \otimes I_{3\times3} &&\in \mathbb{R}^{3N_p \times 3N_p} \,,
\\
&W &&:= \mr{diag}\left(\frac{\omega_{1}}{N_p}, \dots \frac{\omega_{N_p}}{N_p}\right) &&\in \mathbb{R}^{N_p \times N_p} \,,
\\
&\bar W &&:= W \otimes I_{3\times3} &&\in \mathbb{R}^{3N_p \times 3N_p} \,,
\\
&\mathbb{L}^n &&:= \left(\Lambda^n_{ijk}(\etab_1),\dots,\Lambda^n_{ijk}(\etab_{N_p})\right) \qquad \left(n \in \{0,3\}\right)&&\in \mathbb{R}^{N^n \times N_p} 
\\
&\mathbb{L}^n_\mu (\etab) &&:= \left(\vec \Lambda^n_{\mu,ijk}(\boldsymbol{\eta}_1), \dots, \vec \Lambda^n_{\mu,ijk}(\boldsymbol{\eta}_{N_p})\right) \qquad \left(n \in \{1,2\},\, \mu \in \{1,2,3\}\right)&&\in \mathbb{R}^{N^n_\mu \times N_p} \,,
\\
&\mathbb{L}^n &&:=
\begin{bmatrix}
\mathbb{L}^n_1(\boldsymbol{\eta}) & 0 & 0 \\
0 & \mathbb{L}^n_2(\boldsymbol{\eta}) & 0 \\
0 & 0 & \mathbb{L}^n_3(\boldsymbol{\eta}) \\
\end{bmatrix} \qquad \left(n \in \{1,2\}\right)
&&\in \mathbb{R}^{N^n \times 3N_p} \,,
\\
&\frac{1}{\bar{\sg}} &&:= \mr{diag}\left(\frac{1}{\sqrt{g(\etab_1)}}, \dots , \frac{1}{\sqrt{g(\etab_{N_p})}}\right) \otimes I_{3\times3} &&\in \mathbb{R}^{3N_p \times 3N_p} \,,
\\
&\frac{1}{\bar{B}^{*3}_\parallel} &&:= \mr{diag}\left(\frac{1}{\hat{B}^{*3}_\parallel(\etab_1)},\dots,\frac{1}{\hat{B}^{*3}_\parallel(\etab_{N_p})}\right) \otimes I_{3\times3} &&\in \mathbb{R}^{3N_p \times 3N_p} \,,
\\
&\left(\bar 1- \frac{\bar B_\parallel}{\bar B^{*}_\parallel}\right) &&:= \text{diag}\left(1- \frac{\hat B_\parallel(\boldsymbol{\eta}_1)}{\hat B^{*}_\parallel(\boldsymbol{\eta}_1)}, \cdots, 1- \frac{\hat B_\parallel(\boldsymbol{\eta}_{N_p})}{\hat B^{*}_\parallel(\boldsymbol{\eta}_{N_p})}\right) \otimes I_{3 \times 3} &&\in \mathbb{R}^{3N_p \times 3N_p} \,,
\\
&\bar{B}_0(\Hb) &&:= \left(\hat B_0(\etab_1), \dots, \hat B_0(\etab_{N_p}) \right) &&\in \mathbb{R}^{N_p} \,,
\\
&\bar{B}_\parallel(\Hb, \bb) &&:= \bb^\top \mathcal{P}^\top \mathbb{L}^0 &&\in \mathbb{R}^{N_p} \,,
\\
&\bar{\Bb}_0(\Hb) &&:= \left[\bar{\Bb}_{0,\mu}\right]_{\mu=1,2,3}\,, \quad \bar{\Bb}_{0,\mu} := \left(\hat\Bb^2_{0,\mu} (\etab_1), \dots, \hat\Bb^2_{0,\mu}(\etab_{N_p}) \right) &&\in \mathbb{R}^{3N_p} \,,
\\
&\bar{\Bb}_0^\times(\Hb) &&:= \left[\bar{\Bb}^\times_{0,\mu\nu}\right]_{\mu,\nu=1,2,3}\,, \quad \bar{\Bb}^\times_{0,\mu\nu} := \epsilon_{\mu\alpha\nu} \mr{diag}\left(\hat \Bb^2_{0,\alpha}(\etab_1),\dots,\hat \Bb^2_{0,\alpha}(\etab_{N_p})\right) &&\in \mathbb{R}^{3N_p\times3N_p} \,,
\\
&\overline{\nabla B_0}(\Hb) &&:= \left[\overline{\nabla B_0}_\mu\right]_{\mu=1,2,3}\,, \quad \overline{\nabla B_0}_\mu := \left((\hat \nabla \hat B^0_0)_\mu (\etab_1), \dots, (\hat \nabla \hat B^0_0)_\mu (\etab_{N_p}) \right) &&\in \mathbb{R}^{3N_p} \,,
\\
&\bar{\bb}_0(\Hb) &&:= \left[\bar{\bb}_{0,\mu}\right]_{\mu=1,2,3}\,, \quad \bar{\bb}_{0,\mu} := \left(\hat\bb^1_{0,\mu} (\etab_1), \dots, \hat\bb^1_{0,\mu}(\etab_{N_p}) \right) &&\in \mathbb{R}^{3N_p} \,,
\\
&\bar{\bb}_0^\times(\Hb) &&:= \left[\bar{\bb}^\times_{0,\mu\nu}\right]_{\mu,\nu=1,2,3}\,, \quad \bar{\bb}^\times_{0,\mu\nu} := \epsilon_{\mu\alpha\nu} \mr{diag}\left(\hat \bb^1_{0,\alpha}(\etab_1),\dots,\hat \bb^1_{0,\alpha}(\etab_{N_p})\right) &&\in \mathbb{R}^{3N_p\times3N_p} \,,
\\
&\overline{\nabla \times \bb_0}(\Hb) &&:= \left[\overline{\nabla \times \bb_0}_\mu\right]_{\mu=1,2,3}\,, \quad \overline{\nabla \times \bb_0}_\mu := \left((\hat \nabla \times \hat\bb^1_0)_\mu (\etab_1), \dots, (\hat \nabla \times \hat\bb^1_0)_\mu (\etab_{N_p}) \right) &&\in \mathbb{R}^{3N_p} \,,
\\
&\bar{\Bb}(\Hb, \bb) &&:= \left[\bar{\Bb}_\mu\right]_{\mu=1,2,3}\,, \quad \bar{\Bb}_\mu := \bb_\mu^\top \mathbb{L}^2_\mu(\etab) &&\in \mathbb{R}^{3N_p} \,,
\\
&\overline{\nabla B_\parallel}(\Hb, \bb) &&:= \left[\overline{\nabla B_\parallel}_\mu\right]_{\mu=1,2,3}\,, \quad \overline{\nabla B_\parallel}_\mu(\Hb, \bb) := \bb_\nu^\top \mathcal{P}^\top \mathbb{G}_{\nu,\mu}^\top \mathbb{L}^1_\mu(\etab) &&\in \mathbb{R}^{3N_p} \,,
\\
&\bar{\Bb}^\times(\Hb, \bb) &&:= \left[\bar{\Bb}^\times_{\mu\nu}\right]_{\mu,\nu=1,2,3}\,, \quad \bar{\Bb}^\times_{\mu\nu} := \epsilon_{\mu\alpha\nu}\mr{diag}\left(\bb_\alpha^\top \mathbb{L}^2_\alpha(\etab)\right) &&\in \mathbb{R}^{3N_p\times3N_p} \,,
\\
&\bar \Bb_\mr{tot} (\Hb, \bb) &&:= \bar \Bb_0 (\Hb) + \bar \Bb (\Hb, \bb) &&\in \mathbb{R}^{N_p}\,,
\\[1mm]
&\bar \Bb^\times_\mr{tot} (\Hb, \bb) &&:= \bar \Bb_0^\times (\Hb) + \bar \Bb^\times (\Hb, \bb) &&\in \mathbb{R}^{(3N_p)\times (3N_p)}\,,\,,
\\[1mm]
&\bar{B}_{\parallel \mr{tot}} (\Hb, \bb) &&:= \bar B_0 (\Hb) + \bar B_\parallel (\Hb, \bb) &&\in \mathbb{R}^{3N_p}\,,
\\[1mm]
&\overline{\nabla B_\parallel}_\mr{tot} (\Hb, \bb) &&:= \overline{\nabla B_0} (\Hb) + \overline{\nabla B_\parallel} (\Hb, \bb) &&\in \mathbb{R}^{3N_p}\,.
\end{alignat*}
\endgroup

\bibliographystyle{elsarticle-num} 
\bibliography{reference}

\end{document}

%% file: preamble.tex
\newcommand{\Fb}{\mathbf{F}}

\newcommand{\xb}{\mathbf{x}}

\newcommand{\vb}{\mathbf{v}}
\newcommand{\zb}{\mathbf{z}}

\newcommand{\ub}{\mathbf{u}}

\newcommand{\Eb}{\mathbf{E}}
\newcommand{\Bb}{\mathbf{B}}

\newcommand{\bb}{\mathbf{b}}
\newcommand{\cb}{\mathbf{c}}
\newcommand{\Ab}{\mathbf{A}}

\newcommand{\eb}{\mathbf{e}}

\newcommand{\Ub}{\mathbf{U}}
\newcommand{\Cb}{\mathbf{C}}

\newcommand{\Jb}{\mathbf{J}}
\newcommand{\Zb}{\mathbf{Z}}
\newcommand{\pb}{\mathbf{p}}

\newcommand{\Mb}{\mathbf{M}}

\newcommand{\Vb}{\mathbf{V}}

\newcommand{\Hb}{\mathbf{H}}

\newcommand{\etab}{\boldsymbol{\eta}}

\newcommand{\Ginv}{G^{-1}}

\newcommand{\dx}{\tn d^3 x}
\newcommand{\deta}{\,\tn d^3 \eta}

\newcommand{\dvp}{\tn d v_\parallel}
\newcommand{\dmu}{\tn d \mu}

\newcommand{\mr}[1]{\mathrm{#1}}

\newcommand{\fh}{f_\mr{h}}

\newcommand{\hfh}{\hat{f}^\mr{vol}_\mr{h}}

\newcommand{\hsh}{\hat{s}^\mr{vol}_\mr{h}}

\newcommand{\fho}{f_\mr{h0}}

\newcommand{\Bpa}{B_\parallel^\ast}

\newcommand{\hBpat}{\hat{B}_\parallel^{\ast3}}
\newcommand{\hBst}{\hat{\mathbf{B}}^{\ast2}}
\newcommand{\hEso}{\hat{\mathbf{E}}^{\ast1}}
\newcommand{\hbo}{\hat{\mathbf{b}}_0^{1}}

\newcommand{\qh}{q_\mr{h}}
\newcommand{\mh}{m_\mr{h}}
\newcommand{\ngc}{n_\mr{gc}}
\newcommand{\hngc}{\hat{n}^3_\mr{gc}}
\newcommand{\Jgc}{\Jb_\mr{gc}}
\newcommand{\Mgc}{\Mb_\mr{gc}}
\newcommand{\hJgc}{\hat{\Jb}^2_\mr{gc}}
\newcommand{\hMgc}{\hat{\Mb}^1_\mr{gc}}
\newcommand{\vp}{v_\parallel}

\newcommand{\oosg}{\frac{1}{\sqrt{g}}}
\newcommand{\sg}{\sqrt{g}}

\newcommand{\dt}[1]{\frac{\text{d} #1}{\text{d} t}}
\newcommand{\pdt}[1]{\frac{\partial #1}{\partial t}}

\newcommand{\pder}[2]{\frac{\partial #1}{\partial #2}}

\newcommand{\wt}[1]{\widetilde{#1}}

\let\tn\textnormal